\documentstyle[emulateapj,onecolfloat]{article}

\def\etal{{\rm et~al.\ }}
\def\hmpc{\;h^{-1}{\rm Mpc}}

\def\hmpccc{\;h^{3}{\rm Mpc}^{-3}}
\def\hkpc{h^{-1}{\rm kpc}}
\def\kms{{\rm \;km\;s^{-1}}}

\def\kmsmpc{\kms\;{\rm Mpc}^{-1}}
\def\msun{{\rm M_{\odot}}}
\def\lya{Ly$\alpha$}

\newcommand{\PSbox}[3]{\mbox{\rule{0in}{#3}\includegraphics{#1}\hspace{#2}}}

\begin{document}

\twocolumn[

\title{
High redshift galaxies and the Lyman-alpha forest in a CDM universe
}

\author{
Rupert A.C. Croft$^{1,2}$,
Lars Hernquist$^{1}$,
Volker Springel$^{3}$.
Michael Westover$^{1}$, and
Martin White$^{1,4}$
}

\begin{abstract}
We use a cosmological hydrodynamic simulation of a cold dark matter
universe to investigate theoretically the relationship between high
redshift galaxies and the \lya\ forest at redshift $z=3$.  Galaxies in
the simulation are surrounded by halos of hot gas, which nevertheless
contain enough neutral hydrogen to cause a \lya\ flux decrement, its
strength increasing with galaxy mass.  A comparison with recent
observational data by Adelberger \etal on the \lya\ forest around
galaxies reveals that actual galaxies may have systematically less \lya\
absorption within 1 Mpc of them than our simulated galaxies. In order
to investigate this possibility, we add several simple prescriptions
for galaxy feedback on the IGM to the evolved simulation.  These
include the effect of photoionizing background radiation coming from
galactic sources, galactic winds whose only effect is to deposit thermal energy
into the IGM, and another, kinetic model for winds, which are assumed to
evacuate cavities in the IGM around galaxies. We find that only the
latter is able to produce a large effect, enough to match the
tentative observational data, given the energy available from star
formation in the simulated galaxies. Another intriguing possibility is
that a selection effect is responsible, so that galaxies with low
\lya\ absorption are preferentially included in the sample. This is
also viable, but predicts very different galaxy properties (including
clustering) than the other scenarios.

\end{abstract}
 
]

\footnotetext[1]{Harvard-Smithsonian Center for Astrophysics, 60 Garden Street, Cambridge, MA 02138}
\footnotetext[2]{Physics Department, Carnegie Mellon University,
Pittsburgh, PA 15213; rcroft@cmu.edu}
\footnotetext[3]{Max-Planck-Institut f\"ur Astrophysik, Karl-Schwarzschild- Strasse 1, 85748 Garching, Germany}
\footnotetext[4]{Astronomy Department, University of California,
Berkeley, Berkeley CA} 

\section{Introduction}

The advent of large samples of galaxies with which to probe the
Universe at redshifts as high as 3 or 4 is relatively recent (e.g.,
Steidel \etal 1996, Ouchi \etal 2001). Together with the \lya\ forest,
our main observational probe of the intergalactic medium (IGM), this
data provides a wealth of information on large-scale structure
and galaxies, at a time when the Universe was less than
a third of its current age. Within the last few years, our
theoretical understanding of the IGM at these
redshifts has also increased considerably (e.g., Bi 1993, Cen \etal
1994; Zhang, Anninos, \& Norman 1995; Petitjean, M\"ucket, \& Kates
1995; Hernquist \etal 1996, Hui \& Gnedin 1997). In particular,
hydrodynamical simulations of structure formation can now be run in
volumes large enough to directly simulate the formation of galaxies
(albeit crudely) in their proper cosmological context. Such
simulations have been remarkably successful in matching many
properties of the \lya\ forest, and in explaining the clustering
statistics of galaxies, although the prediction of quantitatively
accurate luminosities of model galaxies has remained a challenging
problem. In this paper, we carry out a numerical study of the relationship
between galaxies and the \lya\ forest in a cold dark matter universe.
We concentrate on the Universe at redshift $z=3$, where recent
observations (Adelberger \etal 2002) have suggested that forming
galaxies may have a profound influence on the IGM in their
environment.

The so-called Lyman-Break technique (e.g., Steidel \& Hamilton 1993)
has become one of the most successful methods to find candidate
galaxies at high redshift. While other samples of high redshift
galaxies exist, such as radiogalaxies (e.g., Rawlings \etal 2001), or
magnitude limited samples of galaxies found in the Hubble Deep Field
(e.g., Phillips \etal 1997), the sheer size of Lyman Break Galaxy
(hereafter LBG) samples makes them most useful for our
study. Observationally, LBGs are known to have high star formation
rates (Adelberger \& Steidel 2000, Pettini \etal 2001), and strong
winds with outflow velocities up to 600$\kms$, or even more (Pettini
\etal 2000, 2001).  Their UV spectra imply that their stellar
populations are between $10^{8}$ and $10^{9}$ yrs old (e.g., Pettini
\etal 2001).  The space density of LBGs in present samples at $z=3$ is
about $1.5 \times 10^{-2} \hmpccc$ (for an Einstein de Sitter
Universe), about the same as present day $L_{*}$ galaxies, for a
magnitude limit of 25.5 (see e.g., Adelberger \etal 2002).  It seems
most likely that LBGs represent the massive end of the galaxy
distribution, given their strong clustering (Giavalisco \etal 1998,
Adelberger \etal 1998, Bagla 1998), although alternative theoretical
scenarios have been suggested (e.g., Kolatt \etal 1999). Semi-analytic
modelling of galaxy formation (e.g., Wechsler \etal 2001) and
numerical hydrodynamic simulations (e.g., Katz \etal 1999) have been
used to study how LBGs might form and cluster in a CDM universe.
While many dark matter halos exist which might host galaxies, 
it is still not certain theoretically which 
 would host those with the properties of LBGs.
 Here, we will explore the possibility
that studying the relationship between galaxies and the surrounding IGM 
through the \lya\ forest may provide some clues.

Neutral hydrogen occupying the space between galaxies causes
absorption features to be seen in quasar spectra. Traditionally, this
absorption has been divided into the \lya\ forest (see Rauch 1998 for
a review), Lyman-limit and damped \lya\ systems (e.g., Wolfe \etal
1986), depending on column density. In the present paper, we shall not
make this distinction, but instead use the term ``forest'' to refer to
all \lya\ absorption (as the true forest, in a volume-weighted sense,
accounts for most of the absorption).  In currently popular theories
of structure formation by gravitational instability, \lya\ forest
absorption arises in a continuous fluctuating medium (see e.g., Bi
1991, Weinberg \etal 1997), where gas traces the dark matter, except
on the smallest scales where pressure forces are important. The
optical depth is proportional to the local neutral hydrogen space
density, as in the uniform IGM case considered by Gunn \& Peterson
(1965). 
As there is structure in the IGM, the \lya\ forest is intimately
linked to galaxy and structure formation as a whole. The IGM, being
 a reservoir of gas for forming galaxies can be studied usefully
using the forest. Galaxies also have the potential 
to disturb the simplest picture we have for the state of the intergalactic
gas, at least in their immediate environs, and this can impact
\lya\ forest observations (Theuns \etal 2001, Ferrara \etal 2000).

In this paper, we analyze a large cosmological hydrodynamic
simulation, which is able to just resolve the Jeans' scale in the
\lya\ forest as well as galaxies with total masses down to a few
$\times 10^9\, \msun$. We will explore the relationship between the
galaxies and the forest, looking at the physical state of gas around
galaxies, how it depends on their mass and other properties, and how
this translates into absorption.  We will also apply simple models to
the final simulation outputs in order to examine what may occur if
supernova feedback is able to act more efficiently and in a more
widespread fashion than allowed by the actual star formation
algorithm used in the simulation.  The same underlying numerical model
has been used by White \etal (2002) to look at the relationship
between galaxies and mass at lower redshift, using gravitational
lensing. Here, we restrict ourselves to redshift $z=3$, where good
observations of both the \lya\ forest and LBGs are available.

Recently, Adelberger \etal (2002, hereafter A02) have completed a
comprehensive observational study of the relationship between LBGs and
the \lya\ forest in the spectra of 8 QSOs at redshifts $z>3$. The LBGs
(431 in number) were selected by A02 to lie along the line of sight to
these QSOs, so that the physical state of the IGM near LBGs could be
probed. A02 found that regions of high galaxy density (after averaging
on scales of $\sim 10\hmpc$), are associated with regions of high
\lya\ absorption. On smaller scales, however, the IGM immediately
surrounding LBGs appears to have a deficit of absorption. A02 point
out that this latter result is tentative, and may not be statistically
significant if galaxy redshifts are actually known with worse accuracy
than believed. The possible implications are however intriguing; it
appears to imply that some sort of feedback from LBGs may directly
influence the stucture of the diffuse IGM that is responsible for the
\lya\ forest.  Clearly, it is worth exploring this effect
theoretically. A02 examine this possibility in some detail, developing
a model in which superwinds from LBGs can affect the \lya\ forest
absorption at some distance from galaxies. In the present paper, we
use a numerical simulation, which predicts the properties of the IGM
and the sites and masses of forming galaxies in a CDM universe in
order to compare to some of the observational data presented by A02.

The paper is set out as follows. We describe the simulation, how
galaxies are selected and \lya\ forest spectra made in \S2. In \S3, we
investigate the properties of the IGM close to galaxies, and how the
\lya\ forest and galaxies are related in \S4. We describe several
simple ways of adding extra feedback to the simulations in \S5, and
what their effect is on galaxy-\lya\ forest properties. A summary and
discussion is presented in \S6.

\section{Simulation and analysis techniques}

The cosmological simulation we analyse is of a $\Lambda$CDM model, with
$\Omega_{\Lambda}=0.7$, $\Omega_{\rm m}=0.3$ $\Omega_{\rm b}=0.04$,
and a Hubble constant $H_{0}=67 \kmsmpc$. The initial linear power
spectrum is cluster-normalized with a linearly extrapolated amplitude
of $\sigma_{8}=0.9$ at $z=0$. The simulation followed the evolution of
a periodic box of comoving side length $33.5 \hmpc$, using $300^3$
particles to represent the gas, and $300^3$ to represent the
collisionless dark matter (DM). The particle masses are therefore
$1.5 \times 10^{7}\, \msun$ for the baryons and $1\times10^{8}\,
\msun$ for the DM. The simulation was run using the smoothed particle
hydrodynamics (SPH) code {\small GADGET} (Springel \etal 2001) on 32
processors of the PC-cluster at the Harvard-Smithsonian Center for
Astrophysics, using a force resolution of $6\,\hkpc$ comoving. In the
following, we will quote all length scales either in comoving units, or in
terms of Hubble velocity (in ${\rm km\,s^{-1}}$).  Compared to most
previous numerical work on the \lya\ forest, the present simulation
offers substantially larger volume, making it a more representative
realization of the $\Lambda$CDM universe, while simultaneously
providing high force and mass resolution. This makes it well suited
for the study of forming galaxies and their relation to the
surrounding IGM.

Besides gravitational and hydrodynamical interactions, the numerical
simulation follows radiative heating and cooling processes of a
primordial mix of helium and hydrogen, including interaction with a
spatially uniform ultra-violet radiation background (UVBG).  This UVBG
was generated using the spectral shape and 
amplitude given by Haardt \& Madau
(1996), and represents radiation from quasars processed through the
IGM. With the value of
$\Omega_{b}$ that we use in the simulation, using the amplitude
 of the radiation field prescribed by Haardt \& Madau, 
approximately reproduces the
correct level of \lya\ forest absorption (see Dav\'{e} \etal 1999, for
the amplitude needed with a lower $\Omega_{b}$), 
 and this paper, \S2.2).

The simulation code also models star formation using an algorithm in
which individual SPH particles at sufficiently high density for star
formation to occur represent a multiphase fluid of hot gas, cold
molecular clouds, and stars. Mass and energy is exchanged between
these phases in a similar way as in Yepes \etal (1997), crudely
describing the complicated physical processes thought to regulate star
formation in the interstellar medium (McKee \& Ostriker 1977). In
particular, thermal instability is assumed to lead to the formation of
cold clouds out of rapidly cooling gas at high overdensity. These
clouds themselves form the reservoir of baryons available for star
formation. Among the stellar populations formed, massive stars are
taken to explode as supernovae instantaneously, depositing thermal
energy in the ambient medium of the multiphase fluid. As an additional
feedback process, it is assumed that supernovae also evaporate some of
the cold clouds, thereby returning condensed baryons to the hot
phase. This process establishes a self-regulation cycle for star
formation and feedback, which stabilizes the gas in the ISM against
run-away gravitational collapse.  Full details of the algorithm are
described in Springel \& Hernquist (2001) and in Springel \& Hernquist 
(2002, in preparation).

Although the physical processes that lead to the formation of
individual stars or stellar associations occur much below the
resolution of the simulation, the multiphase prescription was designed
to parametrize them in such as way that the star formation rate is not
strongly dependent on resolution. This can happen with simpler
algorithms which convert all dense cold gas into stars, as higher
resolution typically means that more gas can reach a higher density,
causing the star formation rate to rise (see e.g., the tests in
Weinberg \etal 1999). In the multiphase description, the dynamics of
the ISM is instead aproximated with a set of continuum equations that
have well-defined relations between density, star formation rate and
pressure of the ISM, making it easier to control numerical resolution
effects.  We note that the particular version of the multi-phase
approach used in the present study was an early version of the model
of Springel \& Hernquist (2001). It has been recently refined in a
number of ways, adding for example an improved parameterization of the
cloud-evaporation efficiency and a self-consistent treatment of
galactic outflows, but these changes do not affect the analysis of the
present study.

\begin{figure}[t]
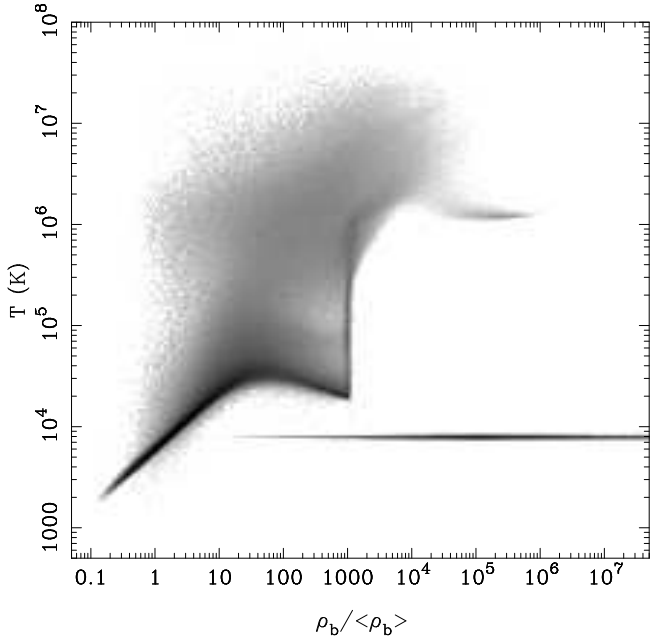

\centering
\PSbox{phase.ps angle=-90 voffset=280 hoffset=-60 vscale=48 hscale=48}
{3.5in}{3.6in} 
\caption[phase]{
Density--temperature phase diagram for the simulation, at redshift $z=3$. 
\label{phase}
}
\end{figure}

In Figure \ref{phase}, we show a phase diagram of temperature
vs. density for particles in the simulation at $z=3$, with a grayscale
weighted by mass. We have plotted separately the two gas components of
the particles which have formed stars. The cold dense cloud component
can be readily seen at the bottom of the plot. Most of the mass in
galaxies is in these clouds, whereas much of the volume of the ISM is
occupied by the hot phase, which has a temperature of close to $10^6$
K and can be seen near the top right of the plot.

Most of the mass in the universe as a whole, however is contained in
the diffuse IGM, at or around the mean density. The locus which
contains this gas is in the bottom left corner, where we can see a
tight relation between temperature and density (see e.g. Katz \etal
1996, Hui \& Gnedin 1997):
\begin{equation}
T=T_{0} \left(\frac{\rho_{b}}{\overline{\rho_{b}}}\right)^{\alpha}.
\label{rhot}
\end{equation}
In this simulation, $T_{0}\simeq 6000$K and $\alpha\simeq0.6$. Both
parameters depend on the reionization history of the gas. If we take a
random particle in the simulation and use its density
$\rho_{b}/\overline{\rho_{b}}$ to predict its temperature $T$ using
Eqn.~(\ref{rhot}), we find that $68\%$ of the mass in the simulation
has an actual $T$ within $50\%$ of this predicted value ($61\%$ within
$25 \%$). This gas is responsible for most of the \lya\ forest
absorption (see Dav\'{e} \etal 1999 for a detailed study).  The
scatter about this relation, while small, is partly due to noise from
the SPH density estimates affecting the cooling rates (Springel \&
Hernquist 2002).

A shocked plume of hot gas occupies space in the top left part of the
plot. This gas has been heated by shocks in IGM filaments and by
falling into potential wells, where it is compressed and passes
through accretion shocks, leading to virialization. Because the
effective temperature of the hot ambient phase of the ISM was fixed at
$\sim 10^6$ K in the multi-phase model used here, some hot ISM gas is
able to escape from small halos and into the IGM, contributing to this
hot plume of gas. The feedback processes modelled by the simulation
therefore have the potential to influence the IGM close to galaxies,
apart from affecting the star formation rate of the collapsed gas
itself. However, we will see below that these effects on the IGM are
rather modest. In \S5 we will add some extra, much more extreme
feedback to the simulations `by hand' using simple models.

We use one other simulation for some tests in \S \ref{lya} below.  It
was run using the same cosmological model, but without star formation.
In this simulation, gas cools and clumps to very high density in
objects which would form galaxies, but the IGM properties are largely
unaffected by the lack of star formation.  The mass resolution of this
simulation is slightly worse ($2\times 224^3$ particles, gas particle
mass $3.6\times 10^{7}\,\msun$), and the box size is the same as for
our multiphase run.

\subsection{Galaxy selection}

Galaxy selection was carried out using a two step algorithm.  We first
identified groups of particles using a friends-of-friends algorithm
(e.g., Davis \etal 1985) with a linking length of 0.15 times the mean
interparticle separation. Embedded in each of the halos thus found,
there can be one or several luminous galaxies, each consisting of
stars, dense and diffuse gas, and dark matter. We thus split the halo
candidates further into gravitationally bound `subhalos' using a
modified version of the algorithm {\small SUBFIND} (Springel et
al. 2001).  Note that due to the large size of the simulation, we had
to fully parallelize both these group identification algorithms in
order to be able to take advantage of the combined memory of several
PC workstations.

After the group identification process has been carried out, we are
left with a list of $\sim 40000$ galaxies with baryonic (gas + star)
masses greater than $M_{b}=2\times 10^{8} \msun$, although only
$33000$ of them already contained independent star particles.  When we
use all galaxies in this list, it should be borne in mind that the
list will not be $100\%$ complete, i.e.~it will not contain all the
galaxies above our mass detection threshold which would have been
identified if we had much better mass resolution. A conservative limit
for completeness is about 64 particles in total (see e.g., Gardner
2001), corresponding to $M_{b}=5 \times 10^{8}\,\msun$ in our case.
For some of our work we will use all the galaxies, although most of
our results will focus on galaxies above a much higher mass threshold,
where completeness is expected to be very good.  For example, with a
mass-cut of $M_{b}=2 \times 10^{10}\, \msun$, there are 405 galaxies
in the volume, so that the galaxies have space density $1.1\times
10^{-2} \hmpccc$, which is similar to the value for local $L_{*}$
galaxies, and somewhat higher than the value for LBGs if the geometry
of a $\Lambda$CDM universe is assumed.

The properties which are available to us after we have constructed our
list of galaxies are their stellar, baryonic and dark matter masses,
their instantaneous star formation rates (SFR), and their positions
and velocities.  The space density of LBGs in the samples of Steidel
and collaborators (see e.g., Adelberger \etal 1999) is $\sim
10^{-3}\,\hmpccc$ for the $\Lambda$CDM geometry.  The simplest
assumption to make is that LBGs correspond to the most massive
galaxies that have formed by redshift $z=3$, and we will do this when
calculating some of our results.  With a lower mass cut of
$M_{b}=10^{11}\, \msun$, there are 30 galaxies in the simulation, so
the observed space density is approximately reproduced.  We can also
select galaxies on the basis of a lower limit in SFR.  As the
simulation does not have sufficient resolution to capture brief
intense starbursts which result from galaxy mergers (e.g., Mihos \&
Hernquist 1996; Hernquist \& Mihos 1995),
 selecting on the basis of SFR effectively chooses
galaxies with high quiescent star formation, which yields a catalog
close to that obtained by choosing on the basis of mass.  We give the
space density of galaxies in the simulation for some values of the
baryonic mass cut in Table~1 and SFR in Table~2.

\begin{table}
\caption{The space density of galaxies, mean galaxy baryonic and total
mass, and mean SFR, in the simulation as a function of lower baryonic
mass cut.}
\begin{center}
\begin{tabular}{ccccc}
 $M_{b}$ cut & $\langle M_{b}\rangle$ & $\langle M_{tot}\rangle$ &
$\langle$ SFR $\rangle$ & Space den.\\
($\msun$) & ($\msun$)& ($\msun$)&($\msun/yr$) &$\hmpccc$\\
$2 \times10^{8}$ & $2.0 \times10^{9}$ & $1.2 \times10^{10}$ & 0.24 & 1.0 \\
$1 \times10^{9}$ & $4.2\times 10^{9}$ & $2.6 \times 10^{10}$ & 0.59 & 0.39 \\
$1 \times10^{10}$& $2.7\times 10^{10}$ & $ 1.7\times 10^{11}$ & 6.0& 0.028 \\
$2 \times10^{10}$ & $4.7\times 10^{10}$ & $3.0 \times 10^{11}$ &12.5& 0.011 \\
$1 \times10^{11}$ & $1.7 \times 10^{11}$&$1.1\times 10^{12}$
 &40.7&$8.0\times 10^{-4}$ \\
\end{tabular}
\end{center}
\end{table}

\begin{table}
\caption{The space density of galaxies, mean galaxy baryonic and total
mass, and mean SFR, in the simulation as a function of lower SFR cut.}
\begin{center}
\begin{tabular}{ccccc}
SFR cut& $\langle M_{b}\rangle$ & $\langle M_{tot}\rangle$ &
$\langle$ SFR $\rangle$ & Space den.\\
($\msun/yr$) &($\msun$)&  ($\msun$)&($\msun/yr$) & $\hmpccc$\\
0.03 & $ 2.9 \times10^{9}  $ & $1.8\times10^{10}$&0.39&0.63 \\
0.1  & $ 4.9\times 10^{9}  $ &$3.0 \times10^{10}$& 0.71& 0.32 \\
1    & $ 3.0 \times 10^{10}$& $1.9 \times 10^{11}$&8.2& 0.020 \\
10   & $ 5.8 \times 10^{10}$ &$3.7 \times 10^{11}$& 21&$5.8 \times 10^{-3}$ \\
\end{tabular}
\end{center}
\end{table}

One additional sample of galaxies which we use is a set of merger
remnants. The selection of these objects, and their properties are
described fully in Westover \etal (2002, in preparation), which also
deals in more detail with the properties of galaxies in this
simulation, and their large scale clustering. Briefly, the merger
remnants are defined to be galaxies which contain particles which were
in at least two separate galaxies at $z=4$. These progenitors were
selected from the $z=4$ simulation output using the same parallel
groupfinder, and were required to contain at least 64 particles each.

\begin{figure}[t]
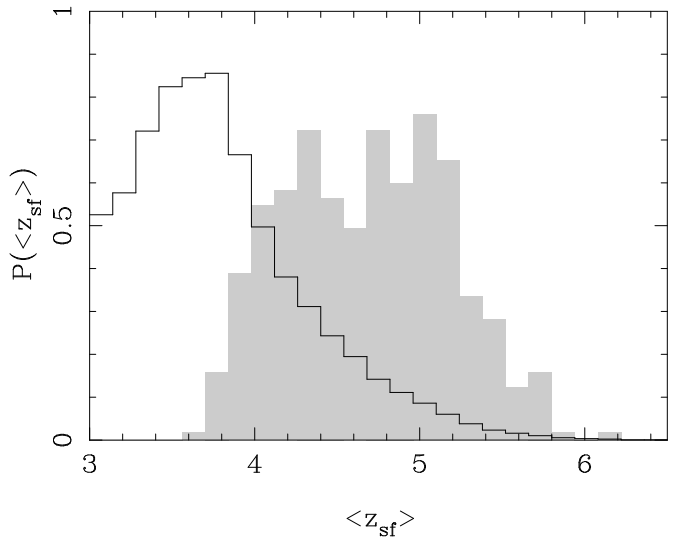

\centering
\PSbox{zsf.ps angle=-90 voffset=280 hoffset=-55 vscale=48 hscale=48}
{3.5in}{3.2in} 
\caption[zsf]{ The probability distribution of the mean redshift of
star formation for galaxies in the simulation, at redshift $z=3$.  The
histogram in outline shows results for all 33000 galaxies in the
simulation volume which contain stars, and the shaded histogram is for
the 405 galaxies with baryonic masses larger than $2 \times 10^{10}\,
\msun$.

\label{zsf}
}
\end{figure}

In the simulation, star formation is occuring vigorously at the $z=3$
output time which we focus on.  Information on the SFRs for galaxies
of different masses can be read from Table 1. Later on, when
implementing our models for extra feedback, we will need to know how
long star formation has been proceeding for each galaxy.  For each of
the galaxies, we have calculated the mean redshift of star formation,
weighted by stellar mass. A histogram of results is shown in Figure
\ref{zsf}, for two different mass cuts: all galaxies, and galaxies
with $M_{b} > 2\times 10^{10} \msun$.  We can see that galaxies which
are massive at $z=3$ formed most of their stars earlier than the small
galaxies, as we would expect in a hierarchical universe where large
structures are assembled from small ones. The mean redshift of star
formation for the massive galaxies is $z=4.75$, whereas it is $3.75$
for all galaxies.  It is interesting that none of the large galaxies
formed most of their stars within the previous $\Delta z \simeq 0.5$,
or $\sim 0.4$ Gyr. The bulk of winds generated by stars in these
galaxies will thus be relatively old by $z=3$.

\subsection{Lyman-alpha forest spectra}
\label{lya}

\lya\ forest spectra were extracted from the simulation by integrating
through the SPH particle kernels along randomly chosen lines of sight
parallel to the axes of the simulation box.  The neutral hydrogen
density in pixels was convolved with the line-of-sight velocity field
in order to generate realistic spectra.  The procedure, which was
carried out using the software package
{\small TIPSY}\footnote{TIPSY is available for 
download at:\\ http://www-hpcc.astro.washington.edu/tools/tipsy/tipsy.html},
is decribed in more detail in Hernquist \etal (1996).  Owing to its
large size, the simulation data could not be analyzed at once on a
single processor, so that it was necessary to modify the
standard software.  In this section, we present results from 500
spectra taken along each axis. In section 5, where different models
are being tested, we restrict ourselves to analyzing 180 spectra in total
for each model.

The simulation outputs at $z=3$ were used, both for the run with star
formation, and for the smaller simulation without star formation
employed for comparison. The effective UVBG intensity was adjusted
slightly after the spectra were extracted, by multiplying the optical
depths for \lya\ absorption ($\tau \propto 1/I$, where $I$ is the
radiation intensity) with a correction factor.  This was done such
that the mean flux level $\left<F\right>=\left<{\rm e}^{-\tau}\right>$
in the spectra was equal to the observed value.  In the tests
involving the \lya\ forest alone, such as those in this section, we
take the observed value to be $\left<F\right>=0.684$, as measured at
$z=3$ from a sample of Keck HIRES spectra by McDonald \etal\ (2000).
For the simulation with star formation, the mean UVBG necessary to
achieve this was 1.16 times the value actually used when the
simulation was run, and 1.15 times for the run without.

When comparing to the data of A02, we use a different value of the mean flux
$\left<F\right>=0.639$. With this
value we approximately
match the mean flux found by A02 at a separation of $1000 \kms$ from galaxies
(\S4) .
 This value (0.639) is the overall mean flux measured at
$z=3$ by Press, Rybicki \& Schneider (1993). To achieve this value, we
lowered the UVBG to 0.85 times the intensity used when running the
simulation. 
As the Press $\etal$ mean flux differs from the McDonald
$\etal$ value, the level of UVBG radiation needed seems to be
uncertain at least at the $\sim \pm 15\%$ level.
The actual overall mean flux found by A02 is actually higher (0.67). The fact
that we match their data with a lower value is probably due to
finite simulation box size effects (we discuss this further in \S4).

\begin{figure}[t]
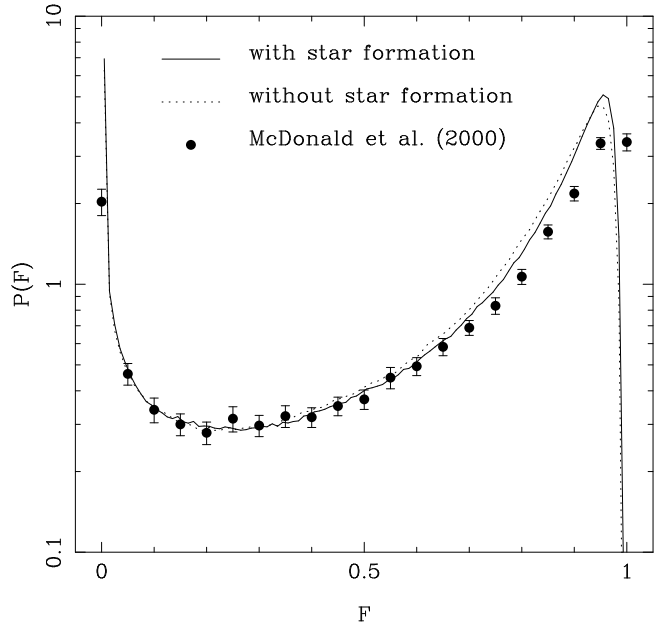

\centering
\PSbox{fpdf.ps angle=-90 voffset=280 hoffset=-60 vscale=48 hscale=48}
{3.5in}{3.6in} 
\caption[fpdf]{The probability distribution of flux for simulations
with and without star formation. The observational results of McDonald
\etal (2000), also at $z=3$, are shown as points with error bars. No
attempt to continuum-fit the simulation spectra was made, so that
systematic errors in fitting the observations account for at least
part of the difference between simulations and observations (see
McDonald \etal 2000).
\label{fpdf}
}
\end{figure}

We have seen that including star formation in the simulation hardly
affects the mean value of \lya\ forest absorption.  The mean UVBG
intensity needed to match the observations differs by less than
$1\%$. In Figure \ref{fpdf}, we show the probability distribution
function of flux values in the spectra for the two simulations and for
the observational results of McDonald \etal (2000). There are
marginal differences between the two theory curves. These may be
attributable to differences in resolution between the two simulations,
as they occur in low absorption pixels of the spectra, which are
likely to be far away from the effects of star formation.  There does
not seem to be any dramatic effect on the absorption from hot gas
which has been forced out of galaxies by feedback, and which can be
seen on the phase diagram (Fig. \ref{phase}). We have plotted the
location of this gas and find it to be confined very close to
galaxies. It has a very small volume filling fraction, not large
enough to affect the \lya\ spectra.  Both simulation curves match the
observations reasonably well, although it is hard to make an accurate
comparison because there are uncertainties involved in estimating the
true quasar continuum (the $F=1$ level) in the observations. McDonald
\etal (2000) have attempted to estimate the possible effect of
continuum fitting errors by modelling the procedure on theoretical
spectra. They find that continuum fitting can change the distribution
of flux values at high $F$ values noticeably, so that our simulations
are likely not inconsistent with the data.

\begin{figure}[t]
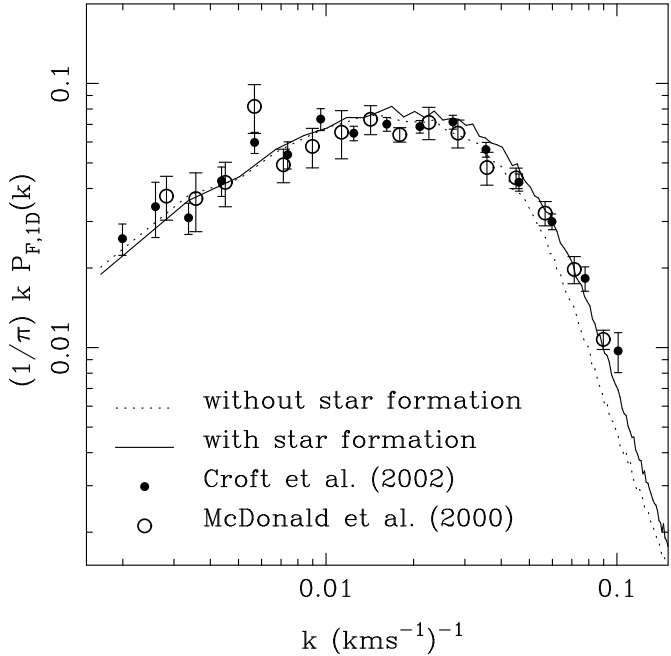

\centering
\PSbox{fpk.ps angle=-90 voffset=320 hoffset=-100 vscale=58 hscale=58}
{3.5in}{3.6in} 
\caption[fpk]{The one-dimensional power spectrum of flux for
simulations with and without star formation (lines). The observational
results of McDonald \etal (2000), and Croft \etal (2002), both also
for $z=3$, are shown as points with error bars.
\label{fpk}
}
\end{figure}

The clustering properties of flux in the spectra can be measured using
the one dimensional power spectrum of the flux,
$P_{F,1D}(k)=\left<\delta^{2}(k)\right>$, where
\begin{equation} 
\delta(k)=\frac{1}{2\pi}\int\delta(x){\rm e}^{-{\rm i}kx} {\rm d}x,
\end{equation}
and which we measure using an FFT. Results are shown in Figure
\ref{fpk}, for the same simulations and observations as in Figure
\ref{fpdf}, as well as the observational data of Croft \etal (2002).
As with the flux probability distribution function, we can see that
differences between the two simulations are generally small. Star
formation and feedback here may be suppressing clustering very
slightly on large scales, but the higher $P_{F,1D}(k)$ on small scales
is probably due to the higher resolution of the star formation
simulation. Both simulations match the clustering of the observational
flux reasonably well, although their amplitude is slightly high. The
turnover on small scales is most sensitive to the temperature of the
IGM (e.g., White \& Croft 2000, Theuns \etal 2000). From
Figure~\ref{phase}, we can see that the temperature at the mean
density is $\sim 6000\,{\rm K}$, which is low compared to
observational Voigt profile fitting results (Schaye \etal 2000,
McDonald \etal 2001).

\section{Galaxy-IGM properties}

\begin{figure*}
\centering
\PSbox{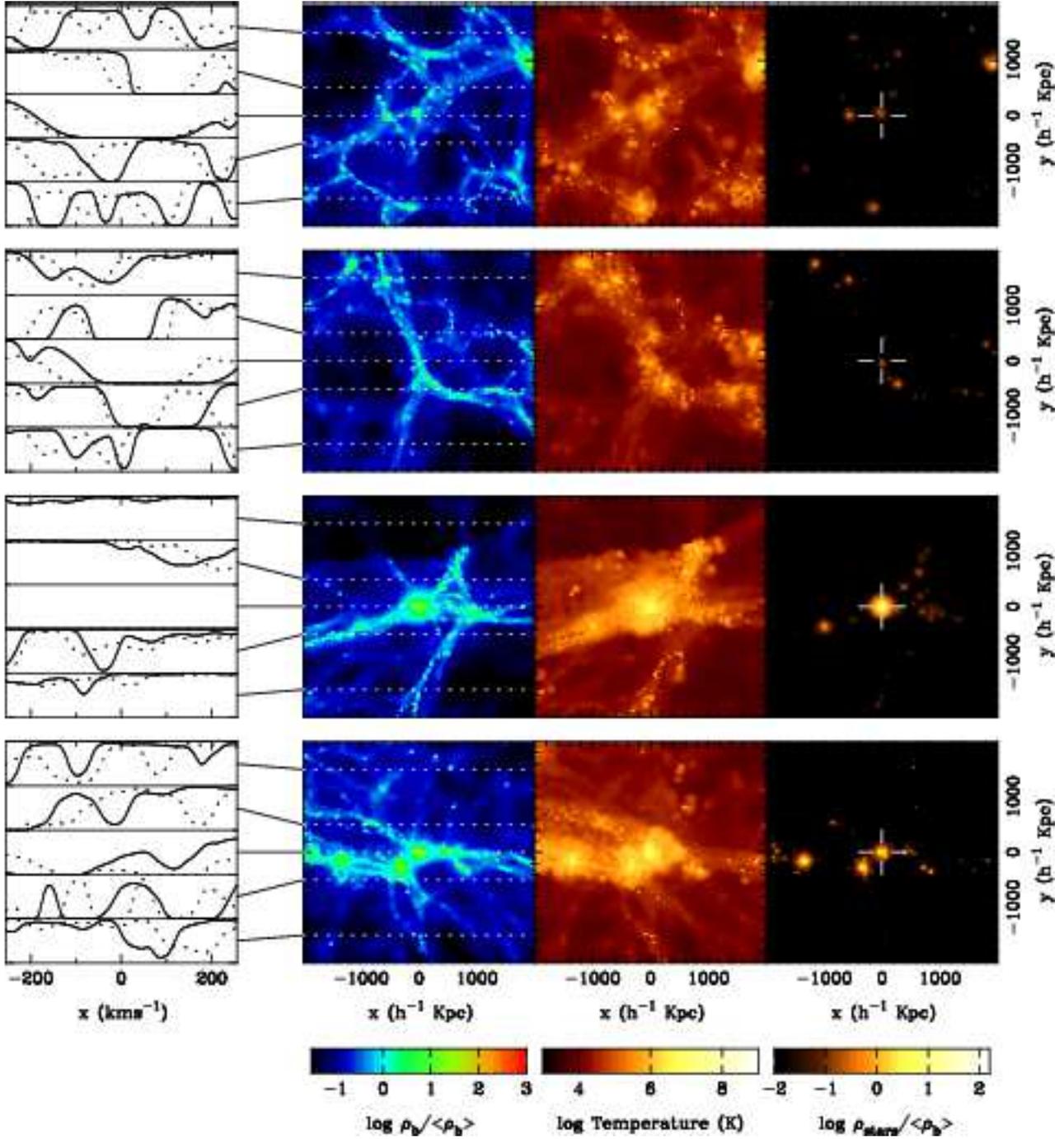 angle=-90 voffset=530 
hoffset=-130 vscale=100 hscale=100}
{3.5in}{7.2in} 
\caption[]{ Rightmost three panels in each row: the gas density, gas
temperature, and stellar density in small regions (the slices are $2
\hmpc$ thick) centered on galaxies in the simulation. The leftmost
panel shows \lya\ absorption spectra extracted along the lines of
sight which are indicated as dashed lines in the density panel. The
solid spectra are the full result, while the dotted spectra have been
calculated with the peculiar velocities set to zero.  The top two rows
show results for two small galaxies (baryonic mass $\simeq 10^{9} \msun$)
and the bottom two rows are for much more massive galaxies (baryonic mass
$\simeq10^{11} \msun$).  }
\label{stamp}
\end{figure*}

We now examine the IGM around galaxies in the simulation. In Figure
\ref{stamp}, we show panels centered on four different galaxies. Two
of the galaxies have baryonic masses $M_{b}\simeq 10^9 \msun$, and the
other two have $M_{b}\simeq 10^{11} \msun$. These examples were chosen
at random from the set of galaxies with masses close to theirs (within
$20\%$). In the three panels on the right of each row, the SPH
smoothing kernels have been used to assign the gas density, the gas
temperature, and the stellar density to a grid of pixels.  Note that
the plots contain particles in a slice of thickness $2 \hmpc$, for
which we show the projected density in units of the mean.

If we first examine the plots of gas density, it appears that the
small galaxies lie along filaments, like strings of beads, whereas the
massive objects tend to be found at filament intersections.  The mass
distribution around the massive galaxies is often strongly anisotropic
outside the central $\sim100\hkpc$. Later, we will be computing
spherically averaged quantities such as the density, temperature and
absorption profiles around galaxies.  It is therefore worth bearing in
mind that the underlying morphology of the gas around galaxies may
influence this.

The panels showing gas temperature are mass-weighted, as we first
assign $\rho_{b} T$ to the pixels before dividing out $\rho_{b}$. The
gas closest to the galaxies has a temperature of about $10^5\,{\rm K}$ for
the small objects and $2 \times 10^{6}\,{\rm K}$ for the large galaxies.
This corresponds to thermal virial velocities of $40\kms$ and 160
$\kms$, respectively.  Although many of the particles are multiphase,
we only count the temperature of the hot component when making the
plot. When we come to calculate the absorption by neutral hydrogen, we
will not include the effect of the cold clouds in galaxies either. As
the clouds have a cross section for absorption themselves, our results
will constitute a lower limit on the absorption. In any case, the
absorption closest to galaxies is subject to several uncertainties,
not least the effect of feedback from winds, and this is what we shall
model in most detail.

In the plots of stellar density in Figure \ref{stamp}, we can see more
clearly where the galaxies actually lie. Both massive galaxies are in
small groups, and even the smaller galaxies have some companions with
similar masses.  

The leftmost panels in each row of Figure \ref{stamp}
show the \lya\ forest absorption along quasar sightlines taken through
the box at three different impact parameters from the galaxies: $0$,
$500$, and $1500\,\hkpc$. We also draw separate curves for spectra
generated along the same sightlines after setting the peculiar
velocities to zero.  By comparing these spectra to the temperature
plots, we can see that there is neutral hydrogen present and causing
absorption even when the gas is hot (the neutral fraction is $\sim
10^{-6}$ in the $\sim 10^6\,{\rm K}$ gas close to the large galaxies).
We notice that neither the shock heating from virialization or the
feedback from SN in the galaxies has cleared out a void in the
absorption.  The absorption becomes noticeably weaker in the
sightlines which have largest impact parameters, even for the smallest
galaxies.  We will see later that there is indeed a substantial
enhancement in absorption within $1-2 \hmpc$ of most galaxies.

When integrated along the portions of spectra plotted, the lines of
sight which have zero impact parameter yield neutral hydrogen column
densities of between $N_{\rm HI}=4 \times 10^{15} {\rm cm}^{-2}$ (the
first galaxy) and $N_{\rm HI}=10^{17} {\rm cm}^{-2}$ (the third
galaxy).  No self-shielding corrections have been applied when
calculating the neutral fraction, which could therefore be higher.
Although the escape fraction for ionizing photons from LBGs has
recently been found to be large (Steidel \etal 2001), we expect that
local photoionizing radiation from stars is unlikely to modify the
absorption properties of this gas (see \S5.1).

\begin{figure}[t]
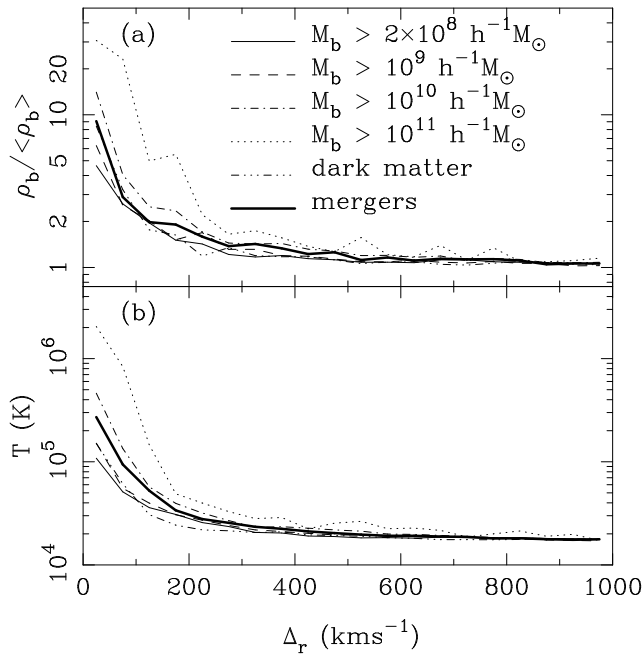

\centering
\PSbox{rhotempavmb.ps angle=-90 voffset=260 hoffset=-50 vscale=44 hscale=44}
{3.5in}{3.6in} 
\caption[rhotempavmb]{ The mean volume-weighted densities (panel [a])
and temperatures (panel [b]), averaged around galaxies in real space
at $z=3$. Results are shown for different lower galaxy baryonic mass
limits, as well as for dark matter particles as centers, and for
galaxies which are merger remnants (see text, \S2.1 for definition).
\label{rhotempavmb}
}
\end{figure}

In order to examine IGM trends with galaxy properties more
quantitatively, we now average the temperature and density around
different sets of galaxies. We use the same sightlines that were used
to make \lya\ forest spectra, and again integrate through the SPH
kernels to find the gas density and temperature in pixels (this time
in real space, without peculiar velocities).  Looping through the list
of galaxies, we measure the distance from the center of each galaxy to
each pixel, in $\kms$.  In our $\Lambda$CDM cosmology, at $z=3$, each
comoving $\hmpc$ corresponds to 112 $\kms$ along the line of sight.
We convert distances across and along the line of sight (sometimes
referred to as $\sigma$ and $\pi$) to $\kms$. The pythagorean
galaxy-pixel distance is the quantity $\Delta_{r}$ plotted on the
$x$-axis of Figure~\ref{rhotempavmb}.  As we calculate the mean
density at each radius by summing the density in pixels and dividing
by the number of pixels, the results are essentially volume-weighted.

In Figure \ref{rhotempavmb}, we show the results for galaxies with
different lower mass limits. There is a general trend whereby higher
mass galaxies are found in regions of higher density, although the
trend with galaxy mass is weak. The profile of density with radius
stays approximately the same in terms of its shape, but the amplitude
varies like $\rho \propto M_{b}^{0.3}$.  The galaxies which are merger
remnants (see \S2.1) have a mean baryonic mass of $ M_{b}\simeq 1.6
\times 10^{10}\, \msun$ and appear to reside in similar density
environments as other galaxies of comparable mass.

The density profile around randomly chosen dark matter particles is
also shown in Figure \ref{rhotempavmb}. Interestingly, although on
large scales the dark matter has lower density and temperature
profiles than the lowest mass galaxies, within $\sim 100\kms$ the
profiles rise significantly. There therefore appears to be some sort
of antibias of galaxies relative to dark matter on the smallest
scales.  We shall see later that this also manifests itself in small
scale clustering measured using the correlation function.  The
temperature profiles all flatten off to the mean volume-weighted
value, which is $17000\,{\rm K}$.

\begin{figure}[t]
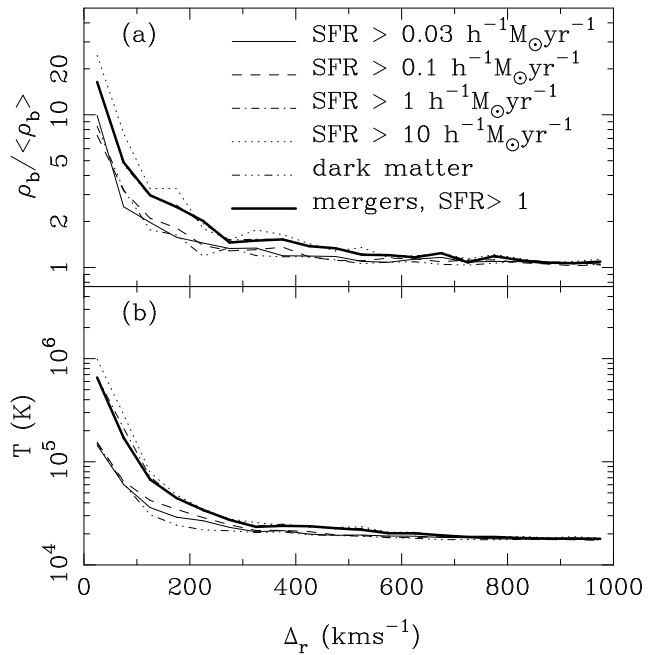

\centering
\PSbox{rhotempavsfr.ps angle=-90 voffset=260 hoffset=-50 vscale=44 hscale=44}
{3.5in}{3.6in} 
\caption[rhotempavsfr]{ The mean volume-weighted densities (panel [a])
and temperatures (panel [b]), averaged around galaxies in real space
at $z=3$. Results are shown for different lower galaxy star formation
rates, as well as for dark matter particles as centers, and for
galaxies which are merger remnants, and which also have star formation
rates above a threshold of $1\,\msun\, {\rm yr}^{-1}$ (see text for
definition).
\label{rhotempavsfr}
}
\end{figure}

If instead we rank the galaxies by their instantaneous SFRs, there is
also a weak trend of increasing density and temperature close to
galaxies, as seen in Figure~\ref{rhotempavsfr}.  In this plot, we show
results for a population of merger remnants slightly different from
that used in Figure~\ref{rhotempavmb}.  Here we take only those merger
remnants with a SFR $> 1\,\msun/{\rm yr}$, which is $25 \%$ of the
merged population. Again, the mergers have similar properties to the
other galaxies with the same SFRs (in fact, the populations largely
overlap, so this is not suprising).

Taking a slightly different approach, we refer to the
density-temperature phase diagram of Figure~\ref{phase}, and to
Equation~(\ref{rhot}).  If we look at galaxies with masses $M_{b}>2
\times10^{10}\,\msun$, $3.4 \%$ of the volume of the universe (and
$13.5 \%$ of the baryonic mass) lies within $0.1-1 \hmpc$ ($11-110
\kms$ in velocity units) of least one of them.  Of this gas, only
$40\%$ by mass lies within $50 \%$ in $T$ of the locus defined by
Equation~(\ref{rhot}). Within $100\, \hkpc$, only $7 \%$ of the gas
lies on the tight $\rho-T$ relation.  The situation for smaller
galaxies follows a similar pattern, but obviously there is much less
space that is at least $1\hmpc$ away from any galaxy.  For all
galaxies in the box, only $33.2 \%$ of the volume, and $12.4 \%$ of
the baryonic mass satisfies this latter condition (being far from all
galaxies), and $99.5 \%$ of the associated gas mass lies on the
$\rho-T$ relation.

We expect the neutral fraction in gas close to galaxies to decline
because of the collisional ionization that occurs at high temperature.
However, the total density of material increases nearby, and the
recombination rate increases, both effects increasing the neutral
density.  We can calculate roughly what we might expect to happen to
the \lya\ absorption if we consider that gas far from galaxies has a
mean temperature of $\sim 20000\,{\rm K}$, and a density $\sim 5-10$
times less on average than gas within $100 \kms$.  The low temperature
gas is almost totally photoionized (see Katz \etal 1996, Croft \etal
1997, Figure 1), and has a neutral fraction of $\sim 4 \times
10^{-6}$. The gas close to massive galaxies is largely collisionally
ionized and has roughly the same neutral fraction, which for
collisional ionization only depends on temperature, and not on
density. The optical depth for \lya\ absorption, being proportional to
the neutral hydrogen density will be roughly $5-10$ times higher
within $100 \kms$ of galaxies.  Note that the recombination rate, and
hence the neutral fraction in the latter case varies like
$T^{-0.7}$. If we wanted to change the temperature so that the \lya\
optical depth would be the same close to galaxies as far away from
them, we would thus need to raise $T$ by a factor of $\sim 10-30$.  We
will explore these issues in \S4 and \S5 in more detail.

\section{Galaxies and the Lyman-alpha forest}

\begin{figure}[t]
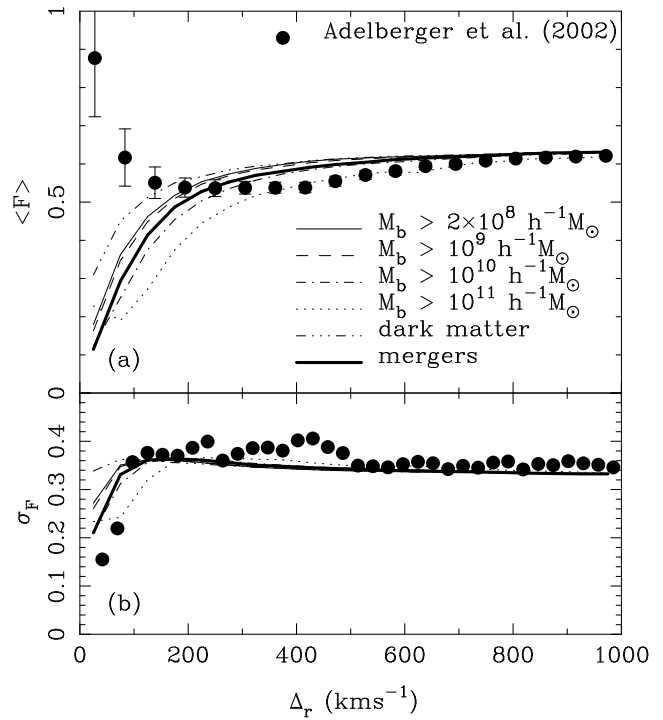

\centering
\PSbox{favsig.ps angle=-90 voffset=300 hoffset=-90 vscale=54 hscale=54}
{3.5in}{4.0in} 
\caption[favsig]{ (a) The mean \lya\ forest flux averaged in pixels at
different distances from galaxies in the simulation (at $z=3$).
Results are shown for different lower galaxy baryonic mass limits, as
well as for dark matter particles as centers, and for galaxies which
are merger remnants (see text for definition).  The observational
results for the LBGs of Adelberger \etal (2002) are shown as points.
(b) The standard deviation of \lya\ forest flux values about the mean
curves shown in panel (a). Again, we show results for pixels at
different distances from galaxies.
\label{favsig}
}
\end{figure}

We take the \lya\ forest spectra extracted from the simulations as
described in \S2.2, and average the flux in pixels at different
distances from galaxies. The results for the mean flux are show in
Figure \ref{favsig}. The quantity shown is defined by
\begin{equation}
\left<F\right> (\Delta_{r})=\frac{1}{N(\Delta_{r})}\sum^{N(\Delta_{r})}_{i=1} F_{i},
\end{equation}
where $N(\Delta_{r})$ is the number of pixels at distance $\Delta_{r}$
($\pm$ a bin width) from a galaxy.  We also plot the standard
deviation of pixel values about the mean curve,
\begin{equation}
\sigma_{F}(\Delta_{r})= \left[ \frac{1}{N(\Delta_{r})}
\sum^{N(\Delta_{r})}_{i=1} [F_{i}-\left<F\right> (\Delta_{r})]^{2}\right]^{\frac{1}{2}}.
\end{equation}
The same lines of sight are used that went into the computation of the
density and temperature profiles in the previous section, except that
the results are now in redshift space.

We saw in \S3 that the temperature and density close to galaxies both
increase for increasing galaxy mass.  In Figure~\ref{favsig}a, where
we show the average flux as a function of $\Delta_{r}$, this
translates to a monotonic relationship between absorption and galaxy
mass. The increase in density is more important than the increase in
collisional ionization that accompanies higher temperatures, and so
more massive galaxies have more associated absorption.  Interestingly,
the absorption profiles around dark matter particles are the weakest,
even though the nearby real space densities around the smallest
galaxies are smaller than for randomly chosen dark matter. This may
have to do with the fact that the \lya\ forest profiles are affected
by infall velocities, which we will explore briefly below.

As with their density profiles, the merger remnants have similar
absorption as the set of all galaxies of the same mass.  Figure
\ref{favsig}a corresponds to a mean profile around many galaxies.
Interpreting this as a single profile around an average galaxy is
misleading (we show below that the dispersion around this profile is
large), even though its shape is suggestive of a Voigt profile, with a
wide damping wing. As we have seen from the density profiles, however,
this is more likely to come from the density enhancement on large
scales that results from clustering.

Recently, observational results have become available for the quantity
$\left< F\right>(\Delta_{r})$, and on Figure \ref{favsig}a we show the
data of Adelberger \etal (2002).  The points were calculated from a
set of 7 fields containing both QSOs and LBG galaxy candidates. 431
LBGs with redshifts were used in the study, and \lya\ forest spectra
were obtained for 8 QSOs with redshifts spanning $3.1 < z <4.1$.  The
error bars on the measured $\left <F\right >(\Delta_{r})$ values
represent the error on the mean obtained from the scatter between
fields.

We can see that the simulations are consistent with the observational
result for scales $\Delta_{r} > 250 \kms$. This is mainly because we
have normalized the UVBG intensity to match the Press \etal 
mean flux measurements, which also reproduces
the A02 mean flux versus distance measurements
on the largest scales we plot.
  The most massive simulation galaxies offer a better fit on
these scales than the others. Because the simulation is not much larger
than the largest scale in Figure \ref{favsig}a, the flux recovers to the
mean more quickly than if larger scale modes were present. Because of this,
to match the A02 data we have used a slightly lower mean overall 
mean flux than A02 found themselves.

On small scales, however, the finite box
size will not cause uncertainties.  The discrepancies we see there are 
real, where the observed mean flux actually rises below
distances of $200 \kms$, signifying a lowering of the mean absorption
as we get closer to galaxies. As this is opposite to what is seen in
the simulations, it is quite surprising, and we will devote a
substantial part of the rest of this paper to studying possible
causes for this.

We note here that the measurement of the absorption profile close to
galaxies is extremely difficult observationally, due to the velocity
differences between different estimates of the galaxy redshift. For
example, the redshift difference between interstellar lines and the
\lya\ emission line often exceeds 750 $\kms$.  According to A02, the
best estimate of the redshift of stars in a galaxy should come from
the nebular lines due to hot gas associated with stars. As these
nebular lines were only available for a minority of galaxies, A02 
had to resort to an estimate of the nebular line 
redshift derived from UV spectral
characteristics for the others. The A02 result shown in Figure
\ref{favsig}a is therefore tentative, as additional redshift
uncertainties might smooth out any absorption dip.  In \S5.2, we will
quantitatively examine an extreme 
version of this effect in the simulation by estimating the
galaxy redshift from material that has the wind velocity.

\begin{figure}[t]
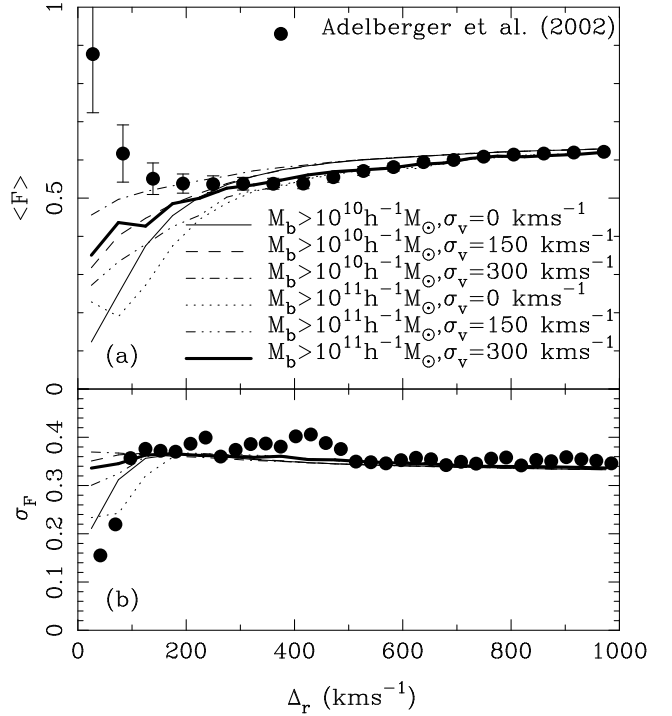

\centering
\PSbox{faverr.ps angle=-90 voffset=300 hoffset=-90 vscale=54 hscale=54}
{3.5in}{4.0in} 
\caption[faverr]{ (a) The mean \lya\ forest flux averaged in pixels at
different distances from galaxies in the simulation (at $z=3$), for
galaxies above two different mass cuts.  The observational results for
the LBGs of Adelberger \etal (2002) are shown as points.  In this
plot, we have added Gaussian velocity errors with a standard deviation
$\sigma_{v}$, as indicated, in order to gauge the effect of
uncertainty in the galaxy redshifts. (b) The standard deviation of
\lya\ forest flux values about the mean curves shown in panel (a).
Again, we show results for pixels at different distances from
galaxies.
\label{faverr}
}
\end{figure}

Here, we will make use of A02's estimates of
the redshift uncertainty. We 
do this by recalculating the results after adding a random
velocity to the redshift of each galaxy, in order to gauge the effect
of the uncertainty. A02 estimate that an optimistic value for the
galaxy redshift uncertainty is given by a Gaussian standard deviation
of $\sigma_{v} \sim 150 \kms$ ($\sigma_{z}\sim 0.002$). A02 also note
that there will in general be a long tail to this distribution, with a
few outliers corresponding to very discrepant results. We have not
modelled this, but restricted ourselves to a value of $\sigma_{v} \sim
150 \kms$, as well as trying $\sigma_{v} \sim 300 \kms$. The results
are shown in Figure \ref{faverr}, for two different mass cuts. We can
see that although the velocity error does smooth out the absorption on
small scales, the effect does not bring the simulations into good
agreement with the data. For the most massive galaxies (which fit the
absorption curves best on large scales) and with the A02 estimate of
$\sigma_{v}$, the effect is not very large.  In the rest of the paper,
we will not convolve our simulation results with errors in this way,
so we should bear in mind that the results on small scales may be
uncertain to the level shown in Figure \ref{faverr}.

The variance around the mean profile is shown in Figure~\ref{favsig}b
(we actually plot the standard deviation $\sigma_{F}$). We might
expect this quantity to be lower around more massive galaxies in the
simulation, as they are more likely to always be accompanied by a
large amount of absorption. We can think of this statistic as probing
the stochasticity of the relationship between \lya\ absorption and
galaxies.  In the plot, the dark matter particles are indeed
accompanied by a larger variance on small scales, and the most massive
galaxies have the least.

The curves rise to a maximum of $\sigma_{F}=0.35$ at $\Delta_{r}\simeq
150 \kms$ and then fall gradually back to a value of $0.32$ at $1000
\kms$ separation. The shape of this function is presumably related to
the fact that within 150 $\kms$ or so of galaxies much of the
absorption is saturated, so that only a small variance in the flux is
possible. On slightly larger scales, the variance of optical depth
about the mean is smaller, but most absorption is in the optically
thin regime, and so the variance in the flux is large. The variance of
the optical depth then decreases further on larger scales.  A02 have
not published this quantity, but the authors were kind enough to
provide preliminary datapoints (Adelberger, private communication).
The observational data follows the same trend, with possibly less
variance on small scales. The exact value of this quantity is
influenced observationally by variations in continuum fitting from
quasar to quasar. These systematic errors will tend to boost
$\sigma_{F}$, so that an accurate assessment of the uncertainties in
$\sigma_{F}$ is difficult, and error bars have therefore not been
plotted.
 We  note that the statistical errors on 
$\sigma_{F}$  are likely to be at least as large than those on 
$\left <F\right >(\Delta_{r})$, so that any inferences we draw
will not be very strong.
We find that the values of $\sigma_{F}$ are slightly lower for the
simulations on large scales, for all cuts in galaxy baryonic mass.  On
small scales, the fact that the observational $\sigma_{F}$ may be
small seems to indicate that \lya\ absorption and therefore the
conditions in the IGM around the LBGs may not vary much from one
galaxy to another.

\begin{figure}[t]
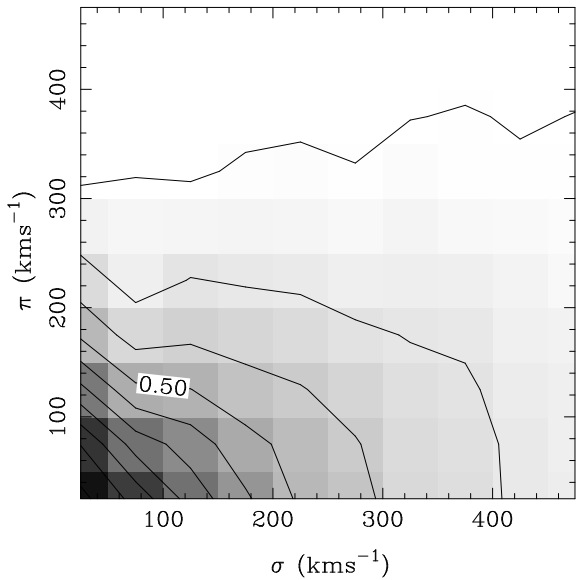

\centering
\PSbox{sigpifav.ps angle=-90 voffset=260 hoffset=-50 vscale=44 hscale=44}
{3.5in}{3.4in} 
\caption[sigpifav]{ The mean value of \lya\ forest flux at different
distances from galaxies (with a lower baryonic mass limit of $2\times
10^{10} \msun$) in the simulation. We show results as a function of
distance across the line of sight ($\sigma$) and along the line of
sight ($\pi$).  The contours are at intervals of 0.05, increasing from
bottom left to top right, and the $F=0.5$ contour is labelled.
\label{sigpifav}
}
\end{figure}

An alternative way to plot the \lya\ absorption around galaxies is to
break up the galaxy's separation into distances across and along the
line of sight ($\sigma$ and $\pi$).  We show a contour plot of
$\left<F\right>(\sigma,\pi)$ in Figure~\ref{sigpifav}.  The effects of
peculiar velocities are clearly visible, including the squashing of
contours on large scales by infall (Kaiser 1987), and the pushing of
absorption contours along the $\pi$ axis due to random virial
velocities close to galaxies.  Such a plot might in principle be used
to study the velocity field of IGM gas around galaxies at high
redshift, or even to measure cosmic geometry (Alcock \& Paczynski
1979), although problems with systematic errors would be considerable.

\section{Simple models for extra feedback}

The star formation model employed in the simulation already
incorporates a simple model for feedback (\S2). For each $\msun$ of
long-lived stars formed, an energy of $7.35 \times 10^{48}$ ergs is
assumed to be released by supernovae from massive stars, and is added
as thermal energy to the ISM, which itself is modelled using
multiphase particles. Because star formation proceeds in a relatively
quiescent manner, the gas is heated up fairly slowly, and dramatic
feedback-related effects do not seem to occur. In particular, the
simulation does not show evidence of the strong winds from high-z
galaxies which are seen observationally (e.g., Pettini \etal 2001).
Ideally, one would like to model star formation more accurately, so
that such features arise naturally. While some progress is being
slowly made in this direction, there are still gaps in our
understanding of the physics of star formation and feedback processes,
which preclude a numerical modelling from first principles
within cosmological volumes at present.
  In the meantime we can explore stronger
feedback in a simplified way by modifying the outputs of our existing
simulation after it has been run.  In this section, we will try out
some very crude, simple models of feedback processes that might be
energetically possible, given the number of SN in each galaxy.  We
will explore the effect of these on the surrounding \lya\ forest
absorption, and how it might differ from that seen in \S4. Also, the
UVBG radiation might be inhomogeneous, and at least partly generated
by galaxies.  We will also briefly examine what consequences this
might have.

The simulation we are using here has already been described in the
previous sections. We are confident that much of the underlying IGM
physics is correct, and also that the cosmological model,
$\Lambda$CDM, is not far from the true one, given that it satisfies
many observational constraints (e.g., Ostriker \& Steinhardt 1995).
The amount of gas which can cool and condense into galaxies is in
principle a reliable prediction of simulations of this kind (Gardner
\etal 2001, and Dav\'{e} \etal 2000), although computations of this
quantity are not without numerical subtleties (Springel \& Hernquist
2002).  However, the simulations probably provide an upper limit,
given that feedback might prevent some condensations from
occuring. While the thermal feedback already included in the
simulations has substantial modelling uncertainty, the exact effects
of the extra feedback
we will add here is even more uncertain. For a start,
we have already used up the available SN energy heating up the IGM and
ISM gas when the simulation was running.  In our simple models, we
will therefore be assuming that this energy input has had little
effect on the IGM, and that we are free to add it again to the
simulation, but in a form where it can have more extreme consequences.

Our approach in this section of the paper is similar in spirit to the
work carried out by Aguirre \etal (2001a,b,c). We will however
be rather less
sophisticated.  Aguirre \etal found that modelling the escape of gas
from the gravitational potential well of galaxies had the most
importance in governing the effects of feedback. We will therefore
treat this in a simplified manner, but we will not explicitly model
the effects of ram pressure, radiation pressure, and the ambient
thermal pressure.  The effects of feedback on the \lya\ forest have
been previously explored by Theuns \etal (2001).  Detailed theoretical
calculations and simulations of the escape of winds from galaxies have
been carried out by, amongst others, Ferrara \etal (2000), Ciardi \&
Ferrara (1997), Efstathiou (2000), MacLow \& Ferrara (1999), and
Murakami \& Babul (1999).

\begin{figure*}[t]
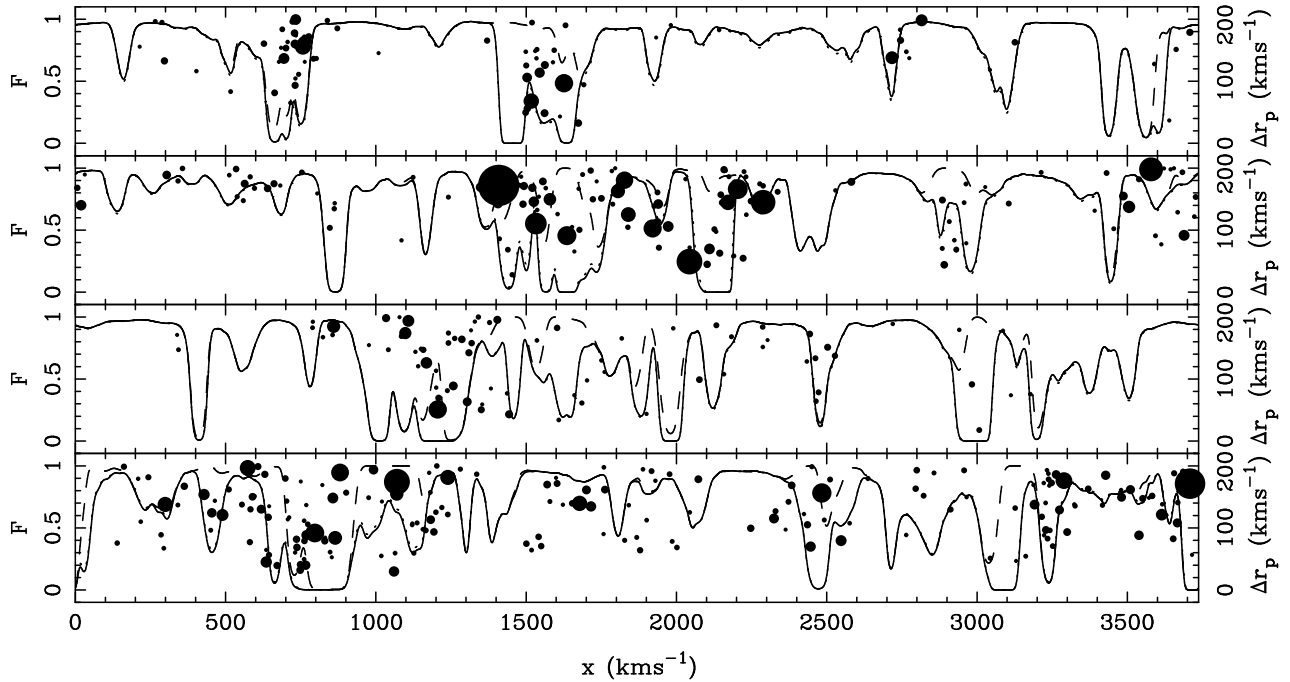

\centering
\PSbox{feedbackspec.ps angle=-90 voffset=400 hoffset=-190 
vscale=80 hscale=80}
{3.5in}{3.7in} 
\caption[]{The four panels show different randomly chosen spectra
through the box.  The fiducial simulation is shown as a solid line,
and two models with different types of feedback from galaxies added
after the simulation was run are also shown.  All spectra were
calculated using the same mean UVBG intensity, but do not have the
same mean absorption.  The dashed line represents a kinetic feedback
model in which IGM gas within a wind radius surrounding each galaxy
(calculated on the basis of the potential well depth and assuming that
10\% of the total supernova energy was available for the winds) was
removed.  The dotted line shows spectra extracted using the same gas
density distribution as in the no extra feedback case, but with an
inhomogeneous UVBG generated by the galaxies themselves.  This line
lies on top of the fiducial line almost everywhere.  The filled points
show the positions of galaxies within $200 \kms$ of the line of
sight. The perpendicular distance from the galaxy to the line of sight
determines its position along the $y$-axis (see scale on the right
$y$-axis). The position along the line of sight in redshift space is
the $x$-coordinate.  The symbol area is proportional in size to the
baryonic mass of the galaxy. The most massive galaxy shown has
$M_{b}=10^{11}\, \msun$.}
\label{feedbackspec}
\end{figure*}

\begin{figure}[t]
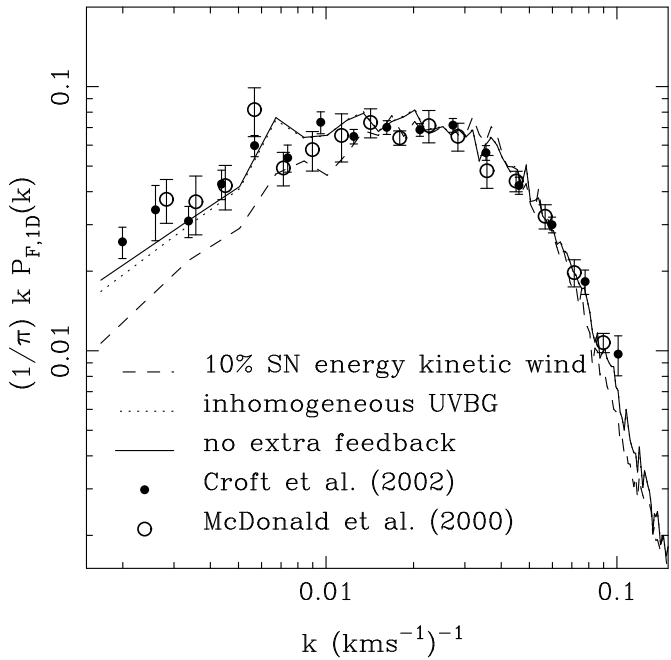

\centering
\PSbox{feedbackfpk.ps angle=-90 voffset=320 hoffset=-100 vscale=58 hscale=58}
{3.5in}{3.8in} 
\caption[feedbackfpk]{The one-dimensional power spectrum of flux for
spectra with no extra feedback (lines), and with two variants of
extreme feedback added to the IGM after the simulation was run (these
are the same as those shown in Figure \ref{feedbackspec}). The dashed
line represents a kinetic feedback model in which IGM gas within a
wind radius surrounding each galaxy was removed.  The dotted line
contains no winds but instead an inhomogenous photoionzing background
radiation field generated by the galaxies themselves.  The
observational results of McDonald \etal (2000), and Croft \etal
(2002), both also for $z=3$, are shown as points with error bars.
\label{feedbackfpk}
}
\end{figure}

\subsection{UV radiation proximity effect}

\label{iuvbg}

As stated in \S2, the simulation was run with a spatially uniform
ionizing background, with a spectrum derived by processing QSO source
radiation through the IGM (Haardt \& Madau 1996). The mean free path
of ionizing photons depends on the amount of neutral hydrogen, which
increases rapidly towards high redshifts. At $z=3$, according to
Haardt \& Madau (1996), the attentuation length $r_{\rm att}$ (over
which the flux decreases by ${\rm e}^{-1}$) is $\simeq 130\,
h^{-1}{\rm Mpc}$ comoving for photons with a wavelength of 912 ${\rm \AA}$,
assuming our $\Lambda$CDM cosmology. The fluctuations of the ionizing
background will depend on this quantity, and on the space density of
sources.  For rare sources such as QSOs, the fluctuations will be
largest, but take place over large scales. Fardal \& Shull (1993), Zuo
(1992), and Croft \etal (1999) showed that even in this case, we
expect little effect on \lya\ forest spectra at $z\sim3$.

There is evidence that a portion (maybe substantial) of the UVBG
arises from starburst galaxies rather than QSOs.  Some flux beyond the
Lyman limit has been seen in spectra of LBGs by Steidel \etal (2001),
indicating that the escape fraction of ionizing photons from these
galaxies may be significant. The implications for the overall level of
the UVBG have been discussed by Haenhelt \etal (2001),  Hui \etal
(2002), and Sokasian \etal (2002).
  If the UVBG was generated by galaxies, fluctuations in its
intensity would occur on smaller scales than with QSOs, but the UVBG
should be smoother overall, as each point in space can ``see'' many
sources. Making use of a model by Kovner \& Rees (1989), Croft \etal
(1999) estimated that sources with the space density of LBGs would
result in the standard deviation of fluctuations being at the $\sim
10^{-3}$ level overall. However, the UVBG close to galaxies would
obviously be of highest intensity, and it is possible that this
``proximity effect'' would decrease the nearby neutral fraction by a
measurable amount.

In order to roughly gauge the effect on the profile of absorption
around galaxies, we have tried replacing our uniform UVBG with point
sources at the positions of galaxies in the simulation. This was done
after the simulation was run, but before spectra were generated.  We
use the optically thin approximation, as the mean free path of photons
is much larger at $z=3$ than our box size. At higher redhifts, this
would not be the case, and simulations that follow reionization
self consistently (e.g., Abel \etal 1999, Gnedin 2000, Ciardi \etal
2001, Sokasian \etal 2001) would become necessary. Here we assume that
the fluctuations in the IGM neutral fraction at $z=3$ do not retain a
strong memory of the earlier history of reionization (see Hui \&
Gnedin 1997, Gnedin 2000).

In our simple model, we take the total flux of ionizing radiation
emitted by each source to be proportional to its instantaneous SFR. We
keep the same UVBG spectral shape as before, which will not affect the
HI neutral fraction, although with the softer spectrum of real
galaxies one might expect other effects, such as less photoheating of
the gas. Radiation from each individual source is allowed to shine on
matter half a box length away, using the periodic boundary conditions,
and then it is cut off sharply.  As this is less than the attentuation
length, we add a uniform UVBG component to account for galaxy sources
farther away.  To calculate this component, we use the fact that the
total intensity of UVBG radiation, and hence the photoionization rate,
at a point in space is proportional to $r_{\rm att}$. This is because
the number of sources within the attentuation volume increases like
$r^{3}_{\rm att}$, but they are dimmed like $r^{-2}_{\rm att}$.  We therefore
add a uniform BG with a photoionization rate
\begin{equation}
\Gamma_{\rm uniform}=\Gamma_{\rm gals}\frac{r_{\rm att}}{0.5\,B},
\end{equation}
where $B$ is the simulation boxsize, and $\Gamma_{\rm gals}$ is the
photoionization rate due to the sources. We estimate $\Gamma_{\rm
gals}$ by Monte-Carlo integration, and take $r_{\rm att}$ to be a
conservative $80 \hmpc$ comoving.  The uniform component of the UVBG
is then $4.8$ times that from galaxies inside the volume. We do not
use any other component for the UVBG, i.e., all of the UVBG is
generated by galaxies.

Some example spectra generated using this inhomogenous UVBG are show
in Figure~\ref{feedbackspec} (the dotted line).  The uniform case is
also shown, and we can see that there is hardly any difference between
them.  As we might expect, the minute observable differences are only
apparent in regions close to galaxies, which are shown on the plot
with their impact parameter relative to the QSO line of sight on the
$y$-axis. The galaxies are mainly concentrated around high absorption
features in the spectra.  The spectra shown as dashed lines in each
panel correspond to a strong kinetic feedback model, which we will
discuss in \S\ref{kfw}.

Overall, owing to their large number, the galaxy sources yield only
small fluctuations in the \lya\ forest absorption.  We have also tried
only using galaxies with high SFRs (e.g. ${\rm SFR}> 10\, \msun\ {\rm
yr}^{-1}$), but it makes little difference.  From our sample of
spectra (180 lines of sight), we find that the overall UVBG intensity
required is the same as in the uniform case. The flux-power spectrum
$\delta_{F,1D}$ is shown in Figure~\ref{feedbackfpk}, compared to that
for the same 180 spectra with a uniform BG. There are only small
differences, rising to $\sim 5\%$ on the largest scales.

\begin{figure}[t]
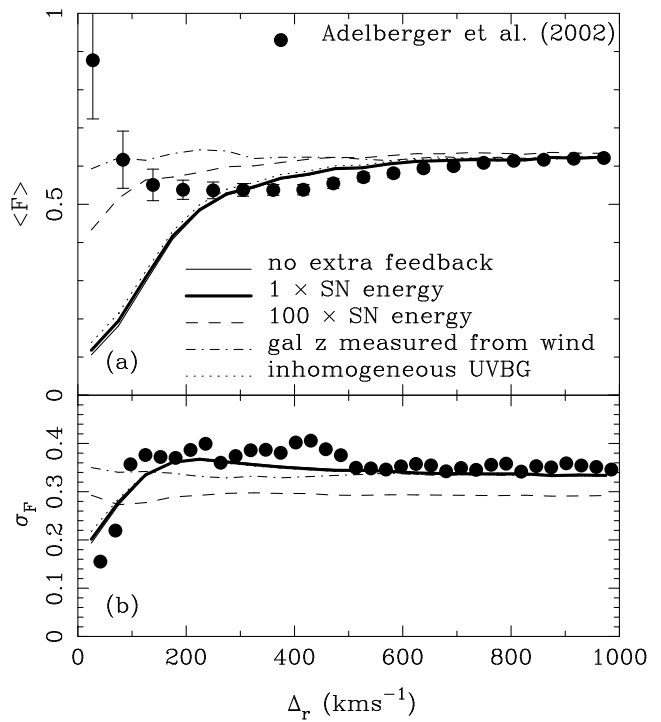

\centering
\PSbox{thermalfavsig.ps angle=-90 voffset=300 hoffset=-90 vscale=54 hscale=54}
{3.5in}{4.2in} 
\caption[thermalfavsig]{ (a) The effect of thermal feedback (for this,
only the temperatures of gas particles are changed, see text,
\S\ref{kfw}) on the mean \lya\ forest flux averaged in pixels at
different distances from galaxies in the simulation (at $z=3$).  All
results are for the flux averaged around galaxies with baryonic masses
$M_{b} > 2\times10^{10}\, \msun$.  The different curves are for
different ways of adding the extra feedback, all computed after the
simulation was completed.  These are a model where $100\%$ of the SN
energy in each galaxy was used to heat the surrounding IGM (thick
solid line), and a similar model, but with 100 times the actual SN
energy available being used (dashed line). The latter is not meant to
represent a realistic model, but was just implemented in order to
yield a noticeable effect.  The dot-dashed line is for results with
$100\%$ of the SN energy used, but with the redshift of the galaxy
taken to be that of the wind (i.e. a difference of $600 \kms$ in this
case).  Finally, the dotted line is for no thermal feedback, but with
the UV background radiation being entirely contributed by the galaxies
in the simulation and hence being inhomogeneous (see \ref{iuvbg}).
The observational results for the LBGs of Adelberger \etal (2002) are
shown as points. In panel (b), we show the standard deviation of \lya\
forest flux values about the mean curves shown in panel (a).
\label{thermalfavsig}
}
\end{figure}

The mean \lya\ forest flux averaged around galaxies with masses
$M_{b}>2 \times10^{10}\, \msun$ is shown in
Figure~\ref{thermalfavsig}, again compared to the same 180 spectra we
computed for the uniform case.  The LBG proximity effect is visible
and has the right sign, but it only changes the mean flux close in by
a maximum of $0.04$.  Modelling the IGM taking into account regions
where there will be an optically thick regime is unlikely to reconcile
the simulation with observations, as such regions would tend to have
more \lya\ forest absorption anyway.  We will describe below the other
lines on Figure~\ref{thermalfavsig}, which are for models of galactic
winds.

\subsection{Thermal feedback from winds}

We will now try distributing SN energy among nearby particles that
might be affected by large-scale winds from galaxies. Given the
theoretical uncertainties involved in predicting the structure of a
wind and its interaction with the IGM, we will just try simple
examples of what might be energetically feasible. One possibility is
that the energy from SN spread by a wind is turned into thermal energy
in the IGM.  Each $\msun$ of stars releases $7.35 \times 10^{48}$
ergs. We have already distributed this energy into the ISM represented
by the multiphase particles. Here we will assume that somehow all of
this energy is able to escape in the form of a wind and instead heats
up the IGM at comparatively large distances. We will however not
remove energy from the galaxy ISM particles, so that we are
effectively adding extra SN energy.

Our model for such ``thermal'' winds is basic. We assume spherical
symmetry, and a wind outflow velocity, $v_{\rm wind}$. For our
fiducial case we set $v_{\rm wind}=600 \kms$, in the same range as the
observational results of Pettini \etal (2000, 2001).  We assume that
the wind starts blowing from each galaxy at the mean redshift of SF of
the particles in the galaxy (see Figure \ref{zsf}), and continues to
$z=3$, without slowing down. In our model for thermal winds, we
therefore do not include the effect of the gravitational potential
well of the galaxy (no matter is moving, and only thermal energy is
propagating outwards). We distribute the thermal energy so that the
energy per unit gas mass in each particle is proportional to
$1/r^{2}$, where $r$ is the distance from the particle to the galaxy
center.  The minimum radius from the center of each galaxy at which
this feedback energy is added is set to be $10\, \hkpc$.  We also
assume that all of this deposited feedback energy remains as thermal
energy at $z=3$, without cooling, in order to achieve the maximum
effect. In the interests of simplicity, we work with only one
simulation output file (at $z=3$), and assume that galaxies do not
move over the wind lifetime.

With $v_{\rm wind}=600 \kms$, we find that the outer radius of the
wind around each galaxy averages $650\, \hkpc$, with a maximum of $2
\hmpc$. As this distance is governed by the mean $z$ of SF, large
galaxies have the largest wind-affected volume around them because
they are older.  With this high velocity and the fact that winds do
not slow down in the model, a very large volume of the Universe is
affected ($60 \%$ is filled by winds). If we reduce $v_{\rm wind}$ to
$300 \kms$, we find a wind filling factor of $18\%$.

\begin{figure}[t]
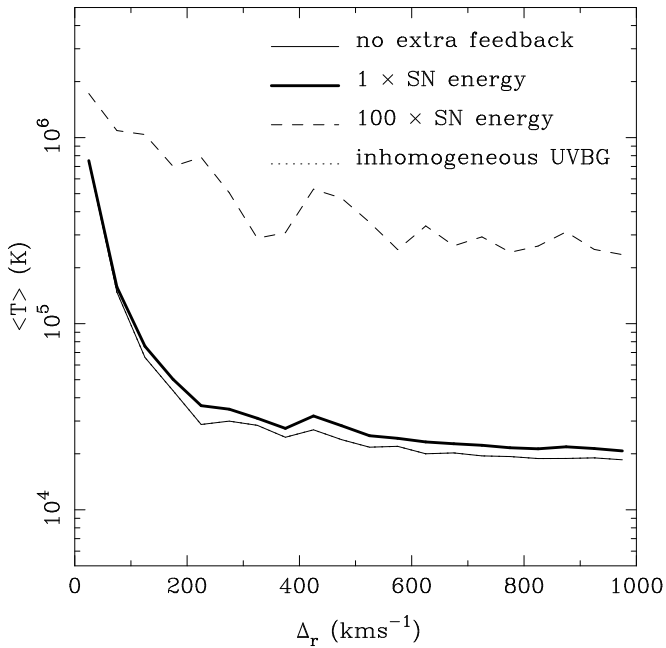

\centering
\PSbox{thermaltemp.ps angle=-90 voffset=310 hoffset=-75 vscale=50 hscale=50}
{4.0in}{4.2in} 
\caption[thermaltemp]{The volume-weighted temperature averaged around
galaxies with with baryonic masses $M_{b} > 2\times 10^{10}\, \msun$,
for different thermal feedback prescriptions (see \S\ref{kfw}, and the
caption for Figure \ref{thermalfavsig} for details). The dotted line
(inhomogeneous UVBG) lies underneath the ``no extra feedback'' line.
\label{thermaltemp}
}
\end{figure}

The effect of the added thermal energy on the mean temperature of the
IGM gas around galaxies is shown in Figure~\ref{thermaltemp}.  In this
plot, we show only results for the 405 most massive galaxies in the
simulation ($M_{b} > 2\times10^{10}\, \msun$).  We can see that with
$100\%$ of the SN energy added, the temperature close in, already high
because of shock-heating, hardly changes compared to the no-wind
case. Further out, the mean temperature rises by $\sim 10\%$. We have
tried varying $v_{\rm wind}$ from $60 \kms$ to $600 \kms$ and changing
the energy deposition law to $1/r^{3}$ but see little difference in
the thermal structure.  In order to see a large effect, we also tried
adding 100 times the available SN energy to the IGM gas, which is
obviously not a realistic model.  The mean temperature then rises to a
few times $10^{5}\,{\rm K}$.  Also on Figure~\ref{thermaltemp} is a
line (hidden under the no feedback line) showing temperature profile
results for the inhomogeneous UVBG model of \S\ref{iuvbg}. Here we
have not added SN feedback energy, and temperature differences (which
are too small to be visible) here only arise from differences in
photoheating rates.

We now calculate \lya\ forest spectra from the treated simulation
output, along 180 randomly chosen lines of sight,  the same that
 were used in the calculation of
 Figure~\ref{thermaltemp}. The mean UVBG
intensity was varied in order to get the correct mean \lya\ forest
optical depth. For the feedback case where $100$\% of the SN energy
was used together with $v_{\rm wind}=600 \kms$, this intensity was the
same as with no extra feedback. For $300 \kms$ winds, the effect of
feedback removed a bit more absorption, so that the UVBG needed was
$1\%$ lower.  For the unphysical model where $100$ times the SN energy
was used, the mean absorption level was drastically affected, and the
UVBG had to be lowered to $48 \%$ of its no-feedback value.

The resulting \lya\ forest absorption profiles around massive galaxies
are shown in Figure~\ref{thermalfavsig}a.  We can see that
distributing the available SN energy as thermal energy has little
effect on the absorption.  It therefore seems likely that to achieve
something with feedback we will need to physically displace gas
surrounding galaxies or disturb its morphology (we will try this
below).  With $100$ times the SN energy used to heat the gas, the
model is closer to the mean \lya\ forest flux results of A02, but this
model is obviously too extreme. Interestingly, the variance of the
flux (Figure \ref{thermalfavsig}b) is lower at large separations in
this unphysical model than in the other cases and in the observations.

Another line on Figure \ref{thermalfavsig}a shows what happens if we
use the redshift of the wind as the redshift of the galaxy when
calculating the galaxy-pixel separation. Basically, in this case we
are subtracting $v_{\rm wind}=600 \kms$ from the galaxy position.
Note that our modelling of this effect is probably too extreme, as the
wind velocity will likely not have the same, high value for all
galaxies. The effect of this offset is as one might imagine from
looking back at Figure \ref{sigpifav}. If one centers the averaging
around the galaxy on a point $600 \kms$ up the $\pi$ axis, the
absorption dip will be missed. Something along these lines may be
occuring in the observations, if the effect of wind velocities on the
measured galaxy redshift has not been completely accounted for.
Evidence against this being the whole story is the fact that,
observationally, the flux rises close to galaxies (although this is
only a $\sim 3 \sigma$ effect), and the variance around the mean
(Figure \ref{sigpifav}b) has the wrong shape for the wind-offset
curve (note, however that this statistic has even larger
 observational uncertanties).

\subsection{Kinetic feedback from winds}
\label{kfw}

We have seen that our thermal wind model, which only changes the
temperature of the IGM does not modify the neutral fraction by a
sufficient amount to affect the \lya\ forest. One can ask about a
different model for feedback, in which winds physically displace gas
in the IGM. In this section we study such ``kinetic'' winds and their
effect on \lya\ spectra.

Once again our model is extremely simple, and we make the assumption
that the winds act in a spherically symmetric fashion around each
galaxy. This time we will use the energy from SN to push baryonic
material away from galaxies.  For each galaxy, we find the
gravitational potential energy necessary to move matter interior to a
shell of radius $r_{\rm shell}$ to that radius:
\begin{equation}
E_{\rm GPE}=G \sum^{N}_{i=1} \left(\sum^{N_{i}}_{j=1} M_{j}\right)
M_{i}\left(\frac{1}{r_{i}}-\frac{1}{r_{\rm shell}}\right)
\end{equation}
where $G$ is the gravitational constant, $N$ is the number of gas
particles interior to $r_{\rm shell}$, and $M_{i}$ is the mass of
particle $i$ at radius $r_{i}$. $N_{i}$ is the number of gas particles
interior to radius $r_{i}$.  Here we ignore the expansion of the
universe over the period of shell propagation. We also assume that all
material within the shell is entrained, obviously the most extreme
scenario possible.

In this simple model, we have a choice when deciding what to do with
the IGM material affected by winds. The easiest scheme to carry out,
and the most drastic, is to make all material inside the outer shell
of the wind disappear. This will result in the evacuation of cavities
in the IGM around each galaxy. As well as this scheme, we also try a
second one, in which the particles are kept with the same SPH
densities and temperatures, but placed at the shell radius.  In order
to avoid a shell profile which is very discontinuous, we smooth out
the shell by adding a random radial displacement to each particle's
position, drawn from a Gaussian distribution with a dispersion of 10\%
of the shell radius.  In carrying out this procedure, we go through
the list of galaxies in a random order, one by one. Because of this,
 some gas particles are moved twice or more,
allowing them to end up close to galaxies they had
previously been moved away from.

As with the thermal wind scenario, the volume of space affected by
winds can potentially be very large. Allowing $100 \%$ of the
available SN energy to move matter with kinetic winds results in a
mean shell radius of $670\, \hkpc$.  This is quite close to the mean
wind propagation radius for galaxies in the thermal wind case, where
winds moved at $600 \kms$ without slowing down.  For $100 \%$ of the
SN energy in kinetic winds we find that $62\%$ of the volume is
affected by winds. Reducing the energy allowed to drive kinetic winds
to $10\%$ results in a wind filling fraction of $28\%$.  Even with
only $1 \%$ of the SN energy in kinetic winds, we still have $10\%$ of
the volume filled by winds.  We also tried the case where only the 400
most massive galaxies in the box are assumed to be able to drive
winds. This resulted in a wind filling fraction of $8\%$ in the $10\%$
SN energy case.

\begin{figure}[t]
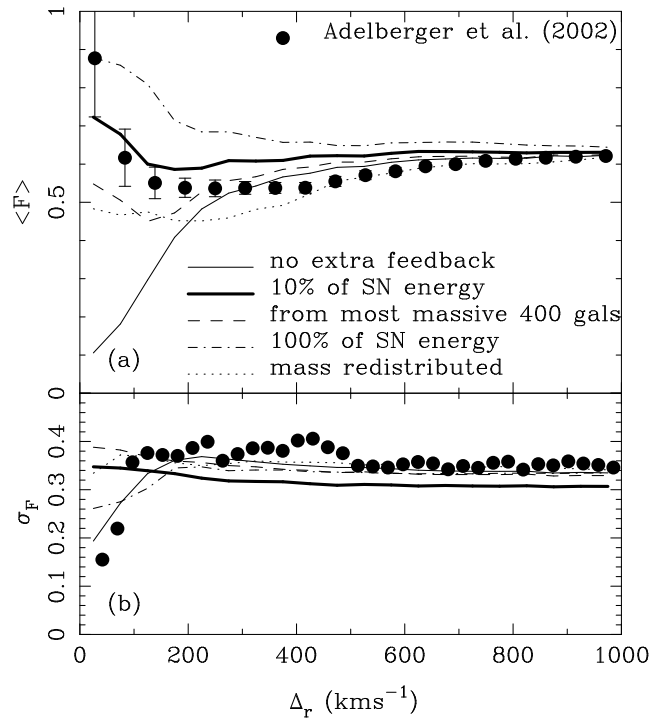

\centering
\PSbox{kineticfavsig.ps angle=-90 voffset=300 hoffset=-90 vscale=54 hscale=54}
{3.5in}{4.2in} 
\caption[kineticfavsig]{ (a) The effect of kinetic feedback (see text,
\S\ref{kfw}) on the mean \lya\ forest flux averaged in pixels at
different distances from galaxies in the simulation, at $z=3$.  All
results are for the flux averaged around galaxies with baryonic masses
$M_{b} > 2\times10^{10}\, \msun$.  The different curves are for
different ways of adding the extra feedback (see \S\ref{kfw} for
details). These are a model where $10\%$ of the available SN energy in
each galaxy was used to evacuate a cavity in the IGM (thick solid
line), and a similar model, but with only SN in the most massive 400
galaxies contributing to the feedback (dashed line) and the fraction
of SN energy increased to $100\%$. Finally, the dotted line shows
results when the mass in the cavity is redistributed in an arbitrary
fashion rather than being eliminated.  The observational results for
the LBGs of Adelberger \etal (2002) are shown as points. In panel (b),
we show the standard deviation of \lya\ forest flux values about the
mean curves shown in panel (a).
\label{kineticfavsig}
}
\end{figure}

We can clearly see the effect of kinetic feedback on individual
features in the spectra if we refer back to Figure~\ref{feedbackspec}.
Here we have plotted the case where winds evacuate gas completely.
Strong lines can be affected, and even if the galaxy involved is
small, the absorption features are completely wiped out when the
impact parameter is small enough.

Our results for the mean \lya\ flux averaged around galaxies are shown
in Figure \ref{kineticfavsig}. As with Figure~\ref{thermalfavsig}, we
only average around galaxies with baryonic masses $M_{b} >
2\times10^{10}\, \msun$.  We can see that, as we might expect,
eliminating the mass within the shells rather than redistributing it
has the largest effect.  If this is done using $100\%$ of the SN
energy (top curve), then there is much too little absorption compared
to the observational data.  With $10\%$ of the energy, there is still
too little, but the results are more reasonable. However, the variance
in the flux on small scales (bottom panel) does not match in this
case.

If the matter is redistributed by winds, this can result in extra
absorption at large ($> 200\kms$) distances from galaxies. In the
particular case shown here, the redistribution results in a better
match to observations on large scales than the case with no feedback.
Of course, the mass redistribution within the feedback scheme is ad
hoc, so that not too much should be read into this. It does however
indicate that it is at least energetically possible for galactic winds
to influence the absorption at some distance from galaxies, in both a
positive and negative way. Mass moved by winds can increase the
overall mean \lya\ absorption, by moving material away from saturated
regions. This could have important implications for the mean UVBG
required to produce the observed level of absorption.

In the case plotted in Figure \ref{feedbackspec} ($10 \%$ SN energy in
winds), the UVBG must be changed by a factor 0.50 with respect to the
no feedback case in order to match the Press et al.~(1993) optical
depth.  If instead of eliminating the material within the wind radius,
we redistribute it, we find that the UVBG intensity must be increased
by 1.57 to match the Press et al.~(1993) results. The effect of
feedback in our simulations on the mean UVBG required is therefore
uncertain. Whether the UVBG must be higher or lower depends on the
absorption lines produced by winds and by the gas moved by them (see
e.g., Theuns \etal 2001, Rauch \etal 2001, for the effects of lines
from wind-blown shells).  Our two kinetic feedback schemes are
extreme, and are likely to bracket the possibilities. For comparison,
our thermal-only feedback model (\S5.2), which does not change the
absorption by much, only requires a UVBG differing by $\sim 2\%$.

We have also calculated the power spectrum of the \lya\ forest spectra
for the kinetic wind model. The results are shown in
Figure~\ref{feedbackfpk}. Unlike the inhomogenous UVBG model, the
kinetic wind feedback is able to change the flux power spectrum
noticeably. Interestingly, this occurs on large scales rather than on
small ones. Theuns \etal (2001) have noted that the distribution
function pixel values of flux in simulations matches that from observed spectra
very well, and concluded that this might be an indication that
feedback is not very important.  This argument may also apply to the
power spectrum. On the other hand, it may be an indication that the
flux power spectrum would be different without feedback, and the
agreement is just a coincidence.  If this were the case, then the
recovery of the power spectrum of matter fluctuations from the \lya\
forest flux power spectrum (e.g., Croft \etal 2002) could be affected.

To summarize this section, among the three different methods we have
tried (UVBG proximity effect, thermal winds, and kinetic winds), only
kinetic winds can potentially have important consequences.
If SN energy in the form of winds
is able to escape from galaxies, and sweep up shells of material (not
just heating it it up), then the absorption properties of the IGM
within $\sim 1 \hmpc$ of galaxies can be significantly altered.  The
apparent deficit of absorption close to Lyman-break galaxies observed
tentatively by A02 could be due to such winds, as suggested by those
authors. We have seen in this section that such a scenario is
energetically plausible in a $\Lambda$CDM universe.

\subsection{Absorption and selection effects}

Rather than LBGs acting on their environments, it is possible that
instead the environment acts on the selection of LBGs themselves.  For
example, let us suppose that the density in dust in the IGM close to
galaxies is linked to the strength of \lya\ absorption.  Galaxies with
little dust obscuration will preferentially be selected to appear in a
magnitude limited sample. These galaxies will then have little \lya\
absorption close to them. If such a selection effect is strong enough,
then it could conceivably account for the deficit of absorption close
to galaxies seen by A02.  In this case, the galaxies in the LBG sample
may also have different clustering properties to the galaxy population
as a whole.

We explore such a scenario in our simulation by taking the set of
\lya\ spectra (with no extra feedback), and for each galaxy looking at
the mean absorption within $100\kms$. We define the absorption as the
mean flux averaged over pixels with $\Delta_{r} <100 \kms$. Ranking
all the galaxies by their nearby absorption, we pick the 400 with the
least, in order to roughly match the space density of LBGs. The mean
baryonic mass of these galaxies is $M_{b}=9\times 10^{8}\, \msun$, half
the value for all galaxies in the box ($1.9 \times 10^{9}\, \msun$).

\begin{figure}[t]
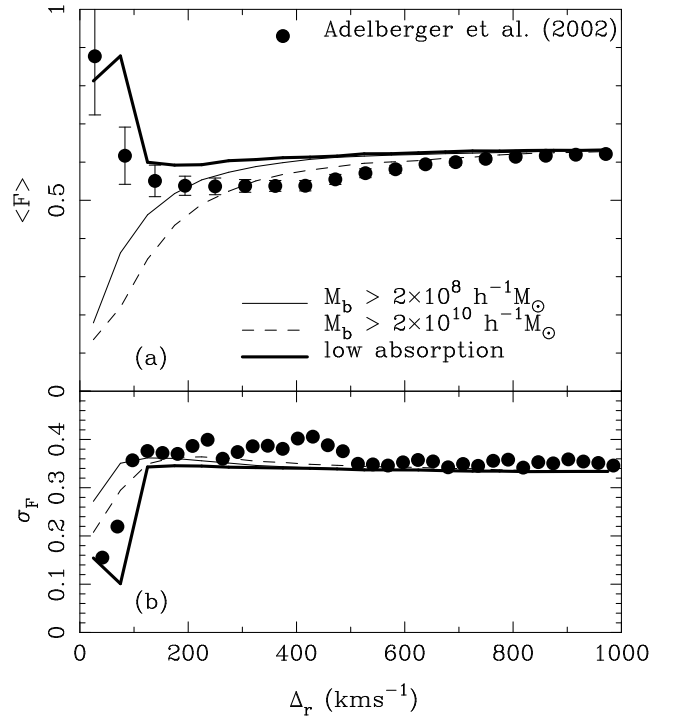

\centering
\PSbox{favsignoabs.ps angle=-90 voffset=300 hoffset=-90 vscale=54 hscale=54}
{3.5in}{4.2in} 
\caption[favsignoabs]{ (a) The mean \lya\ forest flux averaged in
pixels at different distances from galaxies in the simulation, at
$z=3$.  (b) The standard deviation of \lya\ forest flux values about
the mean curves shown in panel (a).  Results are shown for different
lower galaxy baryonic mass limits, as well as for the 400 galaxies
with the least \lya\ absorption within $100\kms$.
\label{favsignoabs}
}
\end{figure}

The galaxy properties will be examined further below. For now, we look
at their absorption profiles. The mean and variance of the \lya\
forest flux around galaxies is shown in Figure~\ref{favsignoabs}.  We
show results for high mass galaxies (the $\sim 400$ most massive
galaxies in the volume), all galaxies, and for the 400 galaxies with
lowest absorption. A sharp rise in the flux on small scales is evident
for the low absorption galaxies.  This means that there does exist a
population of galaxies in the simulation with almost no absorption
within an $\hmpc$ or so.  This is quite surprising, given that we
might expect high absorption to be ubiquitous close to galaxies, given
the absence of strong feedback. The curve for these low absorption
galaxies joins the curve for all galaxies at around $400\kms$. The
panel for the variance in the flux shows that it is lowest for the
low-absorption population. As we picked the galaxies to have
particular values of the flux (low ones), this was obviously going to
occur.  It is still interesting, however, as it offers a way of
differentiating this selection bias model from the models with
feedback, which have higher variance on small scales (see e.g., Figure
\ref{kineticfavsig}.)

\begin{figure}[b]
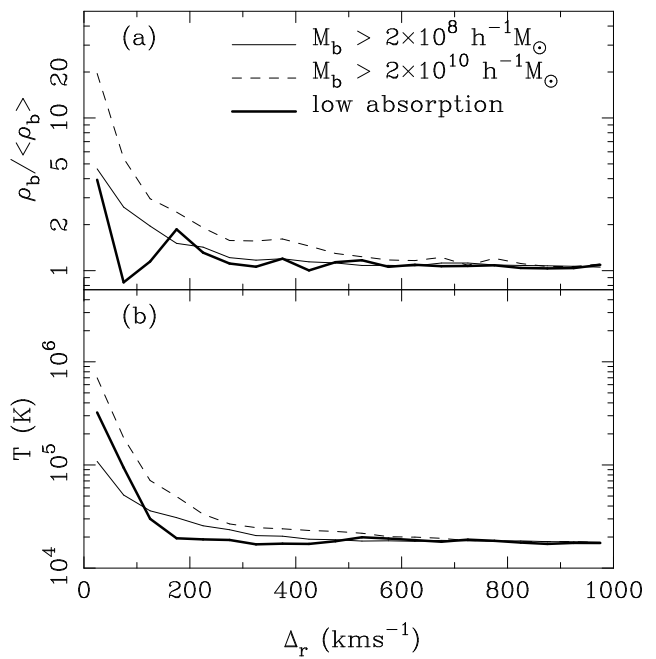

\centering
\PSbox{rhotempavnoabs.ps angle=-90 voffset=260 hoffset=-50 vscale=44 hscale=44}
{3.5in}{3.8in} 
\caption[rhotempavnoabs]{ The mean volume-weighted densities (panel
[a]) and temperatures (panel [b]) averaged around galaxies in real
space, at $z=3$. Results are shown for different lower galaxy baryonic
mass limits, as well as for the 400 galaxies with the least \lya\
absorption within $100\kms$.
\label{rhotempavnoabs}
}
\end{figure}

One can ask why the galaxies with low absorption have the \lya\ forest
profiles that they do. By picking them out, we are supposedly
modelling the effects of dust acting to bias selection. The dust
density and \lya\ absorption are related to each other in such a model
because both depend on the total gas density. The profile of gas
density around the low absorption galaxies should therefore be
systematically low. If instead we see that the lowered absorption
occurs because of a much higher gas temperature (and therefore lowered
neutral fraction), then we should go back and select the galaxies in
another way (e.g., by their nearby gas density). Figure
\ref{rhotempavnoabs} shows the mean volume weighted densities and
temperatures around galaxies, in the same format as in Figure
\ref{rhotempavmb}. We can see that the gas density is indeed low. For
example, the density at $\Delta_{r}\sim 100\kms$ is about 1/3 of the
value for all galaxies. This should result in an optical depth $\tau
\sim (1/3)^{1.6} \sim0.2$ times the value for all galaxies, if the gas
obeys the relationship between optical depth and mass density seen in
the low density IGM (e.g., Croft \etal 1997).  This is enough to
account for the low absorption seen close to these galaxies. If we
look at the temperature profile, however, we can see that for all but
the closest point, the temperature is lower (as would be the case if
density and temperature are related by $T \propto \rho^{0.6}$). The
closest points are puzzling, as the temperature is higher. The Poisson
error bar on the smallest scale point is $7 \%$, so that it seems to
be significant. However, because the dependence of $T$ on $\tau$ is
not as strong as the effect of $\rho$ (\S3), the density effect
dominates the absorption here.

\begin{figure}[t]
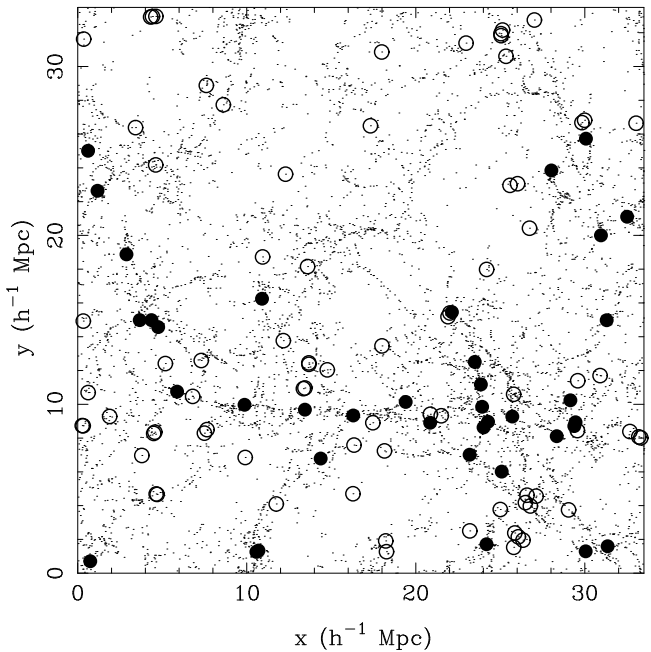

\centering
\PSbox{dots.ps angle=-90 voffset=310 hoffset=-70 vscale=50 hscale=50}
{4.0in}{4.2in} 
\caption[dots]{Galaxies in a slice through the simulation (1/5 of the
box size in the $z$-direction). The small dots represent the total
galaxy population, the open circles the 400 galaxies in the box which
have the least \lya\ absorption within $100\kms$, and the closed
circles the 400 most massive galaxies (in terms of baryonic mass).
\label{dots}
}
\end{figure}

The low absorption galaxies might be expected to have different
clustering properties compared to others. In Figure \ref{dots}, we show the
positions of different galaxy populations in a slice through the
simulation volume. The low absorption galaxies appear to occur
preferentially in low density regions, and are more evenly spread out
than the highly clustered massive galaxies, which prefer filaments.
The morphology of large-scale structure is therefore very
different. The low absorption galaxies do however appear to be quite
clustered on small scales (there are a number of close groups, for
example in the bottom right corner of the plot).

In order to explore this more quantitatively, we have used a
friends-of-friends (FOF, e.g., Davis \etal 1985) groupfinder to find
close groups of galaxies.  If we use a FOF linking length of 0.2 times
the mean intergalaxy separation, we find that the low absorption
galaxies are in groups with a mean number of members equal to 1.2. For
the most massive 400 galaxies, the mean group size is 4.8 members,
while for all galaxies, we find the mean group size to be 1.8
members. From this we infer that the low absorption galaxies do tend
to be isolated. This is in contrast to our first impression derived
from the plot, which could be due to the fact that the space density,
and hence linking length of the three samples is different.

\begin{figure}[b]
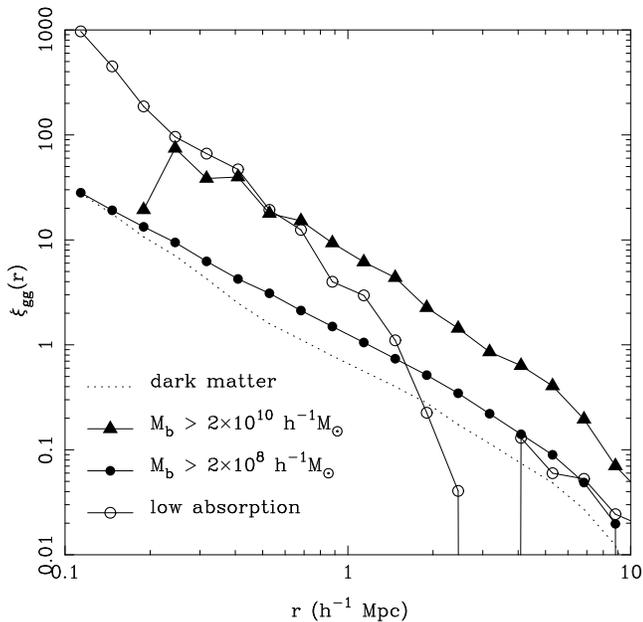

\centering
\PSbox{xi_noabs.ps angle=-90 voffset=280 hoffset=-60 vscale=47 hscale=47}
{4.0in}{3.6in} 
\caption[xi_noabs]{The two-point correlation function $\xi_{\rm gg}(r)$ for
galaxies in the simulation.  We show results for different populations
of galaxies selected as described in the caption for Figure
\ref{dots}, as well as for the dark matter particles.
\label{xi_noabs}
}
\end{figure}

The two point correlation function $\xi_{\rm gg}(r)$ of the different
sets of galaxies is also likely to be different. We plot this quantity
in Figure~\ref{xi_noabs} in real space, without including redshift
space distortions.  Looking at the plot, it is interesting that the
weak absorbers have low clustering on large scales and high on small
scales. The transition between the two regimes occurs close to the
scale which was used to select low absorption ($\sim 1\hmpc$). If we
naively fit a power law, $\xi_{\rm gg}=(r/r_{0})^{-\gamma}$, to the
correlation functions (assuming Poisson error bars), we find that
$r_{0}$ for the weak absorbers is $1.5 \pm 0.05$, and
$\gamma=-2.70\pm0.05$. The equivalent values for all galaxies are
$r_{0}=1.13 \pm 0.03$, and $\gamma=-1.62\pm0.02$. Rather than having
two different power law slopes, however, it is likely, looking at the
plot, that the weak absorber correlation function just has a break at
the selection scale for weak absorption, and then joins the curve for
all galaxies, with the same slope.

The massive galaxies have $r_{0}=2.9\pm0.05$, and
$\gamma=-1.75\pm0.02$.  These values are obviously closer to the
observed values of $\xi_{\rm gg}$ for LBGs than
the results for the low absorption galaxies. For example, Adelberger
\etal (2002) find $r_{0}=3.96 \pm 0.29 \hmpc$, $\gamma=-1.55 \pm
0.15$, assuming $\Lambda$CDM geometry.  The low value of $\xi_{\rm
gg}$ for the low absorption galaxies may therefore be taken as
evidence against the model whereby LBG selection is strongly
influenced by the IGM and dust close to them. Why these galaxies have
strong clustering on the very smallest scales is intriguing. It is
possible that the fact that they have less absorption from IGM gas is
due to the gas having condensed into galaxies nearby, and that it is
this effect which shows up in the $\xi_{\rm gg}$ plot. It is also possible
that the strong small-scale clustering is because the galaxies have
been selected to have similar, rare local environments (Adelberger, private 
communication).

We note that the correlation function of the small galaxy sample is a
very good power law, whereas the dark matter exhibits a break at a
scale of around $0.5 \hmpc$.  The clustering of galaxies in the
simulation is examined in more detail in White \etal (2001).

\section{Summary and discussion}

We have investigated the environment of galaxies which have formed by
redshift $z=3$ in a hydrodynamic simulation of a $\Lambda$CDM
universe. The simulation models star formation in the ISM of galaxies
by using a multiphase prescription, which is expected to be relatively
insensitive to changes in resolution. The star formation rate as a
function of redshift is reasonably close to the observed values, so
that the amount of supernova energy which may affect the IGM should be
fairly realistic.

The galaxies in the simulation are surrounded by gas which is hotter
than in the general IGM, having a temperature of up to $\sim 2\times
10^{6}$ K for the largest galaxies. These large galaxies (with
baryonic masses $M_{b} \sim 10^{10}\,\msun$) tend to lie at the
intersections of several filaments of IGM material, whereas the
smaller galaxies (we resolve galaxies up to 100 times smaller in mass)
lie along filaments. Even though the shock-heated gas surrounding
galaxies has a lower neutral fraction than the cool IGM gas, its high
density means that it is responsible for significant \lya\ absorption.

We take \lya\ forest spectra passing close to the galaxies and average
the \lya\ forest flux as a function of galaxy-pixel distance. In the
simulations, we find that the absorption increases monotonically with
decreasing distance from galaxies.  The absorption is also
systematically stronger for galaxies of greater mass.  On large scales
($> 300 \kms$), for galaxies with baryonic masses $M_{b}> 10^{11}
\msun$, the mean flux level is reasonably consistent with the
observational data of Adelberger \etal (2002). The smallest galaxies
have too little absorption, which might be indicative that LBGs are
more likely to be massive, at least in the context of the model we are
simulating.

We note that Savaglio \etal (2002) have also seen excess absorption in
the \lya\ forest spectrum of a LBG itself. This however was after
averaging over a scale of $> 100 \hmpc$, which is much larger than our
present simulation volume. This is quite a surprising result, and
should be compared to larger simulations.

For galaxy separations less than $300 \kms$, the simulation spectra
continue to have more absorption, whereas the observational data
flatten off, and may even rise on small scales. Some of this
flattening off is probably due to uncertainty in the LBG redshift,
smearing out the absorption feature. However, this could not account
for the rise in flux. If this feature is real, it is possible that
feedback from the galaxies themselves is affecting absorption from the
IGM close by. Observationally, LBGs are known to have large-scale
winds emanating from them (e.g., Pettini \etal 2000).  Star formation
in galaxies modelled with the simulation algorithm used in the present
study is unable to cause such strong feedback. In particular, SN
energy is largely confined to heating the ISM of the multiphase
particles, and not much escapes into the IGM. In the future, we plan
to carry out studies of the \lya\ forest around simulated galaxies
using techniques for treating kinetic winds escaping from galaxies
self-consistently.

In this paper, we have tried adding simple prescriptions for feedback
to the simulation after it has been run. With this approach, we
explore what is energetically possible, given the energy available
from supernovae in the simulation. We find that if the energy of SN is
converted solely to thermal energy in the IGM close to galaxies, then
the resulting temperature increase is too small to affect the neutral
fraction enough to significantly change the absorption. If, on the
other hand, the SN energy is able to physically move gas, evacuating
cavities in the IGM, then only $10\%$ of this energy suffices to
produce an effect as large as that seen in the observations.  The
radius that the winds are able to reach, pushing material out of the
potential wells surrounding the galaxies, is of the order of $\sim 1
\hmpc$.  If the shells of material moved by the wind are able to
contribute to the absorption, then this can change the absorption
profile in a positive way farther out, by adding more
absorption. Because we do not know how to model this redistribution of
gas directly, the profile of absorption in such a feedback model is
uncertain. This also means that the prospects for using the \lya\
forest close to galaxies as a pointer to which simulation populations
(e.g., low or high mass, or mergers) the observed LBG could correspond
to is not promising.

Because LBGs are relatively low in space density compared to all the
galaxies which form in the simulation, we are not putting very strong
constraints on the volume fraction of the Universe which may have been
affected by winds if only observational data on LBGs is considered.
However, since the statistical properties of the \lya\ forest, such as
its power spectrum and flux probability distribution function are
reproduced quite well without strong feedback, it seems that this
volume fraction cannot be too large (see also Theuns \etal 2001), at
least as long as winds affect the \lya\ forest as strongly as
suggested by the observations for the regions close to LBGs. However,
the \lya\ forest may well `tolerate' quite strong thermal winds, or
weaker kinetic winds than the ones tried here, without destroying the
good agreement seen in models without strong feedback.

An alternative explanation for why LBGs may have low absorption close
to them is provided by assuming that a selection effect is
responsible. If low absorption is associated with low obscuration, for
example by dust, then LBGs with low absorption would preferentially
make it into an observational sample. By taking sets of galaxies in
the simulation with the lowest absorption, we find that this can
approximately mimic the \lya\ forest around the observed
LBGs. However, the properties of these galaxies are very different to
the most massive in the simulation box. In particular, they have
small masses, and are weakly clustered on large scales, which is not
favored observationally.

In conclusion, we have seen that the \lya\ forest and high redshift
galaxies in a CDM cosmology are intimately linked. Galaxies form from
the gas which produces the \lya\ forest, and in turn it is
energetically feasible for star formation in these galaxies to affect
the \lya\ forest.  Observations and simulations are starting to be
able to probe the interactions from both points of view. This promises
to be a fruitful step on the way to understanding galaxy formation.

\bigskip
\acknowledgments 
We thank Kurt Adelberger for useful discussions, comments
on the manuscript, and for providing data
in advance of publication.  This work was supported by NASA
Astrophysical Theory Grants NAG5-3820, NAG5-3922, and NAG5-3111, by
NASA Long-Term Space Astrophysics Grant NAG5-3525, and by the NSF
under grants ASC93-18185, ACI96-19019, and AST-9802568.  The
simulation was performed at the Harvard-CfA Center for Parallel
Astrophysical Computing.

\end{document}